\newif\ifextended
\extendedtrue 

\ifextended
\documentclass[sigconf, nonacm]{acmart}
\else
\documentclass[sigconf]{acmart}
\fi

\usepackage{nicefrac}

\AtBeginDocument{%
  }

\ifextended
\else

\copyrightyear{2026}
\acmYear{2026}
\setcopyright{cc}
\setcctype{by}
\acmConference[CCS '26]{Proceedings of the 2026 ACM SIGSAC Conference on Computer and Communications Security}{November 15--19, 2026}{The Hague, Netherlands}
\acmBooktitle{Proceedings of the 2026 ACM SIGSAC Conference on Computer and Communications Security (CCS '26), November 15--19, 2026, The Hague, Netherlands}
\acmDOI{10.1145/3830454.3846608}
\acmISBN{979-8-4007-2871-6/2026/11}

\fi

\usepackage{listings}
\newcommand{\anonymize}[2]{#2}

\ifextended
\newcommand{\refpriorworkcomparison}{\autoref{app:prior_work_comparison}}
\newcommand{\reffulldiscussion}{\autoref{app:full_discussion}}
\newcommand{\refbindzeroanalysis}{\autoref{app:bind-zero-analysis}}
\newcommand{\reftimesidechannel}{\autoref{app:time_side_channel}}
\newcommand{\reftcphijackingflow}{\autoref{app:tcp-hijacking-flow6_4}}
\newcommand{\refreveng}{\autoref{app:reverse-engineering}}
\else
\newcommand{\refpriorworkcomparison}{the extended paper \cite[Appendix D]{extended-paper}}
\newcommand{\reffulldiscussion}{the extended paper \cite[Appendix F]{extended-paper}}
\newcommand{\refbindzeroanalysis}{the extended paper \cite[Appendix E]{extended-paper}}
\newcommand{\reftimesidechannel}{the extended paper \cite[Appendix F.1]{extended-paper}}
\newcommand{\reftcphijackingflow}{the extended paper \cite[Appendix G]{extended-paper}}
\newcommand{\refreveng}{the extended paper \cite[Appendix H]{extended-paper}}
\fi

\begin{document}

\ifextended
\title[]{\texorpdfstring{Cross User/App Network Attacks --- Hijacking TCP Connections and DNS Cache Poisoning via a Malicious User/App \\ (Extended Version) \\ ~\newline \huge{(This is an extended version of the ACM CCS 2026 paper by the same title, \\ \hspace{2.7cm} DOI 10.1145/3830454.3846608)} ~\newline}{Cross User/App Network Attacks --- Hijacking TCP Connections and DNS Cache Poisoning via a Malicious User/App (Extended Version)}}
\fi
\author{Tamir Shahar}
\orcid{0009-0006-0220-1705}
\affiliation{
  \institution{Hebrew University}
  \city{Jerusalem}
  \country{Israel}}
\email{tamir.shahar1@mail.huji.ac.il}

\author{Amit Klein}
\orcid{0000-0002-8024-8756}
\affiliation{
  \institution{Hebrew University}
  \city{Jerusalem}
  \country{Israel}}
\email{amit.klein@mail.huji.ac.il}


\ifextended
\renewcommand{\shortauthors}{}
\fi

\begin{abstract}

Off-path network attacks against TCP and DNS (over UDP) client--server connections are generally considered impractical nowadays, due to built-in security features in these protocols, e.g. randomized TCP (initial) sequence numbers and randomized UDP source ports, respectively. In this work, we refute this presumption by demonstrating that an unprivileged malicious application running on the client (but practically \emph{off-path}), when combined with a remote \emph{off-path} adversary, can enable powerful network attacks against such connections. We show how such a local--remote collaboration between the malicious application and a remote adversary allows inference of sensitive connection state, including TCP sequence numbers and DNS stub-resolver UDP source ports. 

Our attacks exploit standard socket API calls such as \texttt{bind}, protocol mechanisms such as IP options, and operating system features such as cBPF and procfs to infer the TCP initial sequence number (ISN) and the UDP source port in use by the connection of interest. Specifically, we take advantage of certain properties of the ISN generation algorithm as implemented in major operating systems. 

We demonstrate TCP connection hijacking in Linux, Android, Windows, macOS and iOS, and DNS cache poisoning against Windows, Android and the popular systemd-resolved DNS stub resolver in Linux.

We evaluate our techniques across multiple operating systems and realistic deployment settings, including environments behind port-preserving NAT-integrated routers. 

We disclosed our techniques to Microsoft, Apple, Linux and Google, which led to the release of several patches. 

\end{abstract}




\maketitle

\section{Introduction}
\subsection{Motivation}
The security of network protocols such as TCP and UDP against off-path attacks relies de-facto on hard-to-predict fields such as TCP sequence numbers and UDP source ports. These values are assumed to be inaccessible to attackers who cannot observe network traffic, forming a cornerstone of Internet security. At the same time, modern computing environments routinely execute potentially un-trusted local code. Mobile platforms encourage the installation of third-party applications with minimal privileges, while desktop and server systems commonly support multiple users and processes with varying trust levels. In such settings, the presence of a malicious but unprivileged application on the victim host is not an edge case, but a realistic and widespread scenario. 

Our research considers an attacker model in which such a local application collaborates with a remote off-path adversary. Although the local application remains ``off-path'' in the traditional sense---since the operating system prevents it from sniffing packets or accessing raw traffic---it can still interact with OS-level interfaces and observe aspects of network stack behavior. Critically, these interactions are not sufficiently isolated across applications. To illustrate, consider a malicious application installed on a user's device. Without requiring special permissions, it can monitor local connection behavior and coordinate with a remote adversary. 
This raises the central research question of this work: 

\noindent\fcolorbox{black}{gray!20}{
  \parbox{\dimexpr\linewidth-2\fboxsep-11\fboxrule\relax}{
\textbf{Research Question:} Can ordinary operating-system interfaces and network-stack behavior observations available to an unprivileged application reveal sufficient network connection state to enable practical TCP hijacking and DNS cache poisoning when combined with a remote off-path adversary?
}
}
\\

Our research answers this question in the affirmative. Together, the malicious application and the remote adversary can infer sensitive connection state, such as TCP initial sequence numbers and DNS source ports, and use this information to hijack TCP connections during the handshake phase, or poison DNS responses. This enables practical TCP hijacking and DNS cache poisoning attacks against other applications and users on the same device.
This risk is relevant to apps (and system services) which do not use fully encrypted communications. Particularly, although Android's
default policy is blocking cleartext
traffic by apps (starting from Android 9) \cite{android_network_security_config}, a recent large-scale study
found that 33.74\% of apps explicitly enable HTTP (cleartext) connections, and observed actual HTTP
traffic in 2,790 apps out of 35,000 (8\%), leaving such applications directly exposed to
TCP hijacking and DNS-based redirection attacks~\cite{plain-text-plain-risks}.

Our findings suggest that the attack surface extends beyond the network path to the boundary between applications. Even without packet visibility, unprivileged code can extract enough information from shared OS behavior to collapse the effective entropy of protocol defenses. In this work, we demonstrate how this leakage can be turned into practical, end-to-end attacks.

\subsection{Our contribution}

We make the following contributions:
\begin{itemize}
    \item \textbf{Security analysis of TCP ISN generation.}
    We analyze the security of TCP initial sequence number (ISN) as suggested in RFC 6528 \cite{rfc6528} which is included as requirement in RFC 9293 \cite[Section 3.4.1]{rfc9293}. We studied the implementations of the algorithm across major consumer operating systems, including Linux/Android, macOS/iOS, and Windows. Our analysis focus is the local unprivileged application threat model.  
    \item \textbf{Discovery of side channels for ISN and port leakage.}
    We identify multiple side channels that enable leakage of TCP and UDP state. In particular, we demonstrate ISN leakage via IP options and cBPF, and port inference using \texttt{bind} and \texttt{bind(0)} calls.
    \item \textbf{End-to-end attacks in realistic settings.}
    We demonstrate practical exploitation by implementing TCP connection hijacking and DNS cache poisoning attacks in realistic Internet settings, under the local application threat model.
\end{itemize}

\section{Background}

\subsection{TCP Protocol}
The TCP protocol \cite{rfc793_tcp}
is a fundamental, connection-oriented transport protocol that serves as the underlying transport protocol of the web (HTTP/HTTPS), email (SMTP, POP3), file transfer (FTP), and many other application-layer protocols used in modern networked systems, especially over the Internet. To ensure reliable data delivery and preserve packet ordering, TCP assigns a unique sequence number to each byte in the data stream and uses acknowledgment numbers to confirm receipt of data and track progress in the connection stream. 

Each TCP connection endpoint selects a 32-bit ISN when establishing a connection, which determines the starting point of its sequence number space. A TCP connection is established through a handshake process, during which each side exchanges and acknowledges the other's sequence number. In particular, during the TCP handshake, the peer must acknowledge this value by setting the acknowledgment number to ISN+1, thereby confirming receipt of the initial sequence number. The unpredictability of the ISN provides a degree of protection against off-path attackers attempting to inject packets into an active connection.

If an attacker is able to infer or predict the ISN of a client, together with the other necessary connection parameters (i.e. the TCP 4-tuple consisting of source and destination IP addresses and ports), they can impersonate the server 
during connection establishment and subsequently inject arbitrary packets into the established connection. By exploiting this capability, an attacker can target HTTP connections, forging HTTP responses on behalf of a genuine server, potentially enabling phishing attacks. Likewise, an attacker can target other TCP-based protocols such as SMTP, POP3, FTP, etc.

\subsection{cBPF}
Classic BPF (cBPF) \cite{bpf_1992}
is a lightweight packet filtering mechanism implemented in the Linux kernel, which predates the introduction of the more modern and widely known eBPF. 
While eBPF is more powerful and widely used, it requires elevated privileges that are not available in our threat model (\autoref{sec:threat_model}),\footnote{In standard Linux configurations, unprivileged users are typically prevented from loading eBPF programs by default kernel settings. On Android, although the underlying kernel configuration is permissive with respect to eBPF support, a mandatory access control mechanism (SELinux) enforces policies that prevent unprivileged applications from loading or attaching the required programs.} whereas cBPF can be used by unprivileged applications.
cBPF was originally designed to avoid the overhead of copying every packet to user space by instead executing filtering logic directly inside the kernel. Unlike eBPF, cBPF provides a much simpler virtual machine with a restricted instruction set, consisting of basic operations such as loading data from packet offsets, simple arithmetic and logical operations, and conditional branching. There is no straightforward mechanism for cBPF to communicate information back to user space. However, a key property of cBPF is that it can truncate packets before they are delivered to user space, thereby controlling how many bytes are passed to the socket, which can be used as a leakage channel.

\subsection{IP Options}
The Internet Protocol (IP) \cite{rfc791_ip} 
supports optional header fields, known as IP options, which allow additional control information to be carried within packets. These options are not commonly used in typical Internet traffic, but remain part of the protocol and are supported by end hosts and some network devices. IP options are primarily associated with debugging and measurement functionality. One such option is the Loose Source and Record Route (LSRR) 
option, which allows the sender to specify a sequence of intermediate routers that the packet should traverse. As the packet is forwarded, each router may record its address in the option field, enabling the sender to observe the path taken by the packet. LSRR permits routers to forward the packet through intermediate hops not explicitly listed, as long as the specified nodes are visited in order. Another related option is Strict Source and Record Route (SSRR), 
which requires the packet to follow the exact sequence of specified nodes without deviation. In particular, source routing mechanisms allow the sender to influence the routing path of packets, potentially causing them to traverse specific network locations. While LSRR and SSRR are commonly filtered in many network environments, they may still be supported and forwarded along some network paths. 
An early study found source-routing capability in about 8\% of discovered routers
\cite{heuristics_for_internet}.
\subsection{DNS Protocol}
The Domain Name System (DNS) \cite{rfc1034_dns,rfc1035_dns} 
is a fundamental auxiliary protocol that provides name resolution services, translating human-readable domain names into IP addresses.
It is used by a wide range of Internet applications, including web browsing (HTTP/HTTPS), email delivery (SMTP), and other services. DNS primarily operates over UDP using a query-response model. A consumer operating system typically runs a stub resolver --- a system-wide service responsible for issuing DNS queries --- which sends requests to a recursive DNS resolver server and receives corresponding responses. 
Each DNS query is identified by a 16-bit transaction identifier (TXID), and responses are matched to queries based on this identifier, as well as the source and destination UDP ports and IP addresses (the UDP 4-tuple). To improve performance and reduce latency, DNS stub resolvers cache responses, allowing subsequent queries for the same domain to be answered locally.

To provide protection against off-path attackers, DNS relies on the unpredictability of the TXID and the use of randomized UDP source ports. These mechanisms drastically increase the difficulty of forging a valid response that will be accepted by the stub resolver.
However, if an attacker is able to infer or predict these values, they can in fact inject forged DNS responses and poison the stub resolver's cache. 
As a result, traffic intended for benign domains can be redirected to attacker-controlled servers, enabling hijacking of web (HTTP) responses and email messages (SMTP), as well as traffic of other application-layer services.

\section{Related Work}

\subsection{Host-Based and OS-Level Side Channels}
Several works have explored how low-privilege code can extract sensitive information from operating-system behavior. In particular, Zhang et al.~\cite{leakage_on_ios} showed that even on platforms such as iOS, where traditional interfaces such as procfs are unavailable, unprivileged applications can infer cross-app activity using coarse system statistics exposed through system APIs. Their study demonstrates that OS-level abstractions may unintentionally leak information despite strong isolation mechanisms. In contrast, rather than inferring user or application activity, we extract network-related state that can be used to support off-path injection attacks.

A handful of studies have established the power of collaborative
attacks in which unprivileged code on the victim cooperates with a remote off-path adversary. In particular, Qian et al.~\cite{collaborative_tcp_squence_number_inference} showed that attacker-controlled unprivileged code running on the victim can observe differences in connection behavior and thereby validate TCP sequence-number guesses via side channels, enabling off-path
injection. However, their techniques for Android, iOS, and macOS have already been mitigated. For example, their Android attack relies on global counters available via \texttt{/proc/net}, which has been restricted for unprivileged applications on modern Android systems since Android~10 \cite{android10_privacy_changes}, thereby limiting its practical applicability. 
In addition, we demonstrate TCP attacks on Windows and introduce DNS cache-poisoning attacks, which were not considered in their work. We develop new primitives for inferring ISNs that do not rely on the previous global-counter approach.

Chen et al.~\cite{static_detectino_of_packet_injection_vulnerabilities} proposed static analysis techniques to identify packet injection vulnerabilities by uncovering attacker-controlled implicit information leaks; however, their work focuses on detection rather than demonstrating practical exploitation. Furthermore, for their attacks to be effective against Linux, they require the application of the techniques in Qian et al. (see above).

At the operating-system level, Spreitzer et al.~\cite{procHarvester} showed that unprivileged applications can exploit side channels exposed through the \texttt{/proc} filesystem to infer system activity and cross-process behavior; in response, access to procfs resources has been continuously restricted since Android M (Android 6), and access to global interrupt statistics was removed in Android O (Android 8).

\subsection{Off-Path TCP Injection and State Inference}

A substantial body of work has demonstrated that TCP connections can be hijacked by off-path adversaries through side-channel leakage of connection state. 
Gilad et al.~\cite{attacking_the_web} showed that globally shared protocol mechanisms, such as IP identifier (IPID) counters, can be exploited to infer sequence numbers and inject data into active TCP sessions; this behavior has since been modified in modern systems. Cao et al.~\cite{global_rate_limit} revealed additional side channels, including the global challenge-ACK rate limit in TCP stacks, which can be used to validate sequence number guesses and enable off-path injection attacks; this attack has since been mitigated \cite{linux-patch-for-challenge-ack}. Feng et al.~\cite{mixed_ipid_assigment} demonstrated that even mixed IPID assignment strategies may still leak sufficient information for TCP exploitation, but this was since fixed \cite{linux-patch-ipid-unsharing}. Wang et al.~\cite{packet_size_side_channel} and Li et al.~\cite{packet_length_leakage} explored cross-layer side channels in wireless networks, showing that packet size and length can leak TCP state and enable injection and hijacking attacks even when traffic is encrypted. These attacks rely on characteristics of wireless environments (e.g., cellular or Wi-Fi networks), whereas our approach does not require such network-specific characteristics. 

Feng et al.~\cite{pmtud_breaks_tcp} showed that side channels can arise from shared network infrastructure, specifically exploiting PMTUD state to infer connection information and enable TCP hijacking. However, their attack relies on an attacker-controlled puppet 
(i.e., a fully controlled machine in the local network) that can observe fine-grained network side-channel information (e.g., packet sizes). In contrast, our approach only requires a regular unprivileged malicious application running on the victim and does not assume such fine-grained visibility.

Yang et al.~\cite{session_manipulation_attacks} demonstrated that VPN connection tracking can also introduce side channels that enable inference of connection state and subsequent injection attacks. Their attacks, however, rely on VPN-specific deployment conditions, whereas our approach does not depend on such assumptions.

Feng et al.~\cite{PMTUD_is_not_Panacea} demonstrated that IP fragmentation can be exploited to inject malicious payloads without requiring sequence number inference. However, their attack relies on the feasibility of hitting the genuine 16-bit IPv4 ID, which does not carry over to the 32-bit IPv6 fragment ID field, 
and is therefore not applicable in IPv6 networks, whereas our approach remains effective in both IPv4 and IPv6.

In contrast to these approaches, which rely on externally observable network behavior, our work considers a complementary setting in which state inference is achieved through \emph{local} interaction with the victim host.

\subsection{DNS Cache Poisoning Attacks}
The fundamental security of DNS against off-path response forging relies on the unpredictability of the DNS TXID field (16 bits) and the UDP source port (14--16 bits) of the DNS query. Additional mechanisms that increase security exist (e.g. DoH/DoT, DNS cookies, and DNSSEC), but they are irrelevant to our threat model (see \autoref{sec:threat_model}). Prior work has shown that these defenses (randomized DNS TXIDs and UDP source ports) can be weakened in practice. In particular, Klein~\cite{cross_layers_attack_klein} showed that weaknesses in kernel-level randomness can allow remote attackers to predict UDP source ports, significantly reducing the effort required for DNS poisoning; however, this issue has since been fixed. 

Alharbi et al.~\cite{collaborative_client_side_dns} and Hay et al.~\cite{dns_posioning_via_port_exhaustion_2011} demonstrated that attackers can defeat UDP source-port randomization by exhausting the available UDP port space on the target system, thereby reducing the entropy of port selection and enabling DNS cache poisoning. 
Notably, Alharbi et al. utilized the exact same collaborative threat model considered in our work, carrying out the attack by combining an unprivileged malicious application on the victim with a remote off-path attacker.
Similarly, Hay et al. achieved port exhaustion by executing a malicious Java applet within the victim's browser to forcefully allocate and consume the available UDP sockets.

However, these approaches effectively create a denial-of-service condition, as other applications relying on ephemeral UDP ports may fail while the ports are exhausted. Similarly, Gierlings et al.~\cite{Isolated_and_exhausted} showed that remote attackers can exhaust UDP source ports via browser-based techniques; 
however, this issue has since been mitigated in modern browsers. In contrast to these approaches, which either predict or manipulate port allocation, our work infers the stub resolver's source port through local observation and interaction with the system, without exhausting the port space or disrupting normal operation. 

Herzberg and Shulman \cite{fragmentation_considered_poisonous} proposed alternative techniques for reducing entropy in DNS defenses, particularly through the use of IP fragmentation. 
While these attacks primarily target recursive resolvers and DNS forwarders, we speculate that similar techniques may also be applicable to stub resolvers. At the same time, current best practices recommend avoiding IP fragmentation in DNS deployments altogether, e.g., by limiting response sizes and rejecting fragmented DNS traffic. Moreover, these attacks target IPv4, whose IP ID field is 16 bits; in IPv6, whose fragment ID is 32-bits, these attacks are ineffective.
In contrast, our attack does not rely on IP fragmentation and is therefore unaffected by these mitigations, and in addition it is effective in IPv6 networks.

A summary table comparing our work and the related literature is provided in \refpriorworkcomparison.

\section{Threat Model}
\label{sec:threat_model}
\paragraph{Common Threat Model.}
We consider an adversary composed of two cooperating entities: (i) an unprivileged malicious application executing on the victim machine, and (ii) an Attacker Remote Node (ARN), which is off-path and therefore cannot observe traffic between the victim and external servers. 
The unprivileged malicious application threat model has been considered in prior work~\cite{leakage_on_ios, collaborative_tcp_squence_number_inference, procHarvester, collaborative_client_side_dns}. 
Due to its unprivileged execution environment, the malicious application cannot access raw sockets or perform low-level network operations, and cannot directly observe network traffic or fine-grained connection state. It is therefore practically \emph{off-path}. 
We assume a port-preserving router in both IPv4 and IPv6 
settings,\footnote{Recent measurement studies show that port preservation is common in practice; for example, 75 out of 92 evaluated Wi-Fi networks (81\%) were found to exhibit port-preserving behavior in IPv4 settings~\cite{port_preservation}. In IPv6, where NAT is typically not used, port preservation is effectively inherent.} and that the victim's externally visible IP address is known to the ARN, as it is revealed during standard client--attacker communication. Under this model, we consider two attack scenarios: TCP connection hijacking and DNS cache poisoning.

\paragraph{Spoofing Capability.} 
We assume that the ARN can craft and send packets with arbitrary headers, including spoofed source IP addresses, for example using raw sockets with elevated privileges. In addition, we assume that the underlying network permits IP spoofing. This assumption is justified by prior measurement studies showing that a non-negligible fraction of networks still permit IP spoofing; for example, measurements indicate that 14.9\% of IPv4 prefixes and 30.5\% of Autonomous Systems (ASes) allow spoofed-source traffic~\cite{souce_address_validation_in_the_internet}. In such networks, spoofed packets can reach the victim and be accepted as part of legitimate communication, enabling both TCP injection and DNS response forgery.

\paragraph{TCP Hijacking Scenario.}
The adversary--malware pair targets TCP connection attempts to a specific service identified by a known destination IP address and port. We consider two deployment settings: (i) the ARN is located outside the victim's local network, and (ii) the ARN is co-located within the victim's local \emph{switched} network, yet remains off-path (this allows us to consider some scenarios as described in \autoref{sec:ip-option-preprocessing}). The attack targets connection establishment (i.e., newly initiated connection). In this setting, the malicious application enables inference of the client's ISN (which is not directly observable to an unprivileged application) and related TCP connection state, allowing the ARN to inject forged packets that impersonate the server and hijack the connection. 

\paragraph{DNS Cache Poisoning Scenario.}
The adversary targets the victim's stub resolver, whose upstream recursive resolver IP address is assumed known. We assume that the malicious application can trigger DNS queries for domains chosen by the attacker. By default, the stub resolvers we attack do not verify DNSSEC signatures, do not use encrypted DNS protocols (DoH/DoT/DoQ), and do not employ DNS cookies. In this setting, the malicious application enables identification of the UDP source port used by the stub resolver, reducing the attack to a brute-force race over the transaction identifier (TXID). By sending forged responses that match the correct port and a guessed TXID, the attacker can inject malicious records into the stub resolver's cache.

\section{ISN Generation Algorithm}
The ISN is a transport-layer (Layer 4) parameter within TCP which is not directly accessible from user space; consequently, our objective is to infer its value using a non-privileged application on the target device. 
Achieving this requires understanding how the ISN is generated. The security of the ISN relies on the use of a secret hashing key in its generation process. However, we show that under our threat model, the manner in which the hash output is used in the ISN calculation significantly undermines this protection. The ISN is derived from a deterministic algorithm as described in RFC 6528 \cite{rfc6528} and RFC 9293 \cite[Section 3.4.1]{rfc9293}. The algorithm is implemented as-is in the Unix-like operating systems we surveyed; the Windows implementation deviates in certain aspects, though it retains similar structural properties.

\subsection{Linux and Android}
\label{sec:linux_isn}
The TCP ISN is computed as: 
\begin{equation}
ISN = H_k(4\text{-tuple}) + t \mod 2^{32}
\end{equation}
where $H_k$ denotes a keyed hash function over the connection 4-tuple (source IP, destination IP, source port, destination port), and $t$ is a timestamp with 64\,ns resolution. 
This construction combines a per-connection deterministic component with a time-dependent offset. The keyed hash incorporates a 
128-bit secret value $k$ generated at boot time, which is not accessible from user space, and ensures that connections with different 4-tuples receive independent base values. 
The rationale behind the time component design is to ensure that, for a fixed 4-tuple and for a duration of about two minutes, sequence numbers of new connections are always larger\footnote{In the $\mod 2^{32}$ cyclic sense.} than those of previous connections using the same 4-tuple, thereby preventing delayed packets belonging to earlier connections from being accepted as valid data, while the keyed hash addendum prevents off-path attackers from predicting the ISN of new connections. However, since for a fixed 4-tuple, the hash component remains constant, if an attacker can observe a single ISN value $ISN_1$ at time $t_1$, they can approximate a future ISN value $ISN_2$ at time $t_2$ (where $t$ has 64\,ns resolution) as:

\begin{equation}
ISN_2 = ISN_1 + (t_2 - t_1) \mod 2^{32}
\end{equation}

This relationship significantly reduces the entropy of the ISN from the attacker's observation point, enabling prediction under certain conditions.

\subsection{macOS and iOS}
Both macOS and iOS have their network stack implemented as part of the XNU kernel. Although XNU differs from the Linux kernel, the XNU ISN generation algorithm is almost identical. In this case, $t$ is a timestamp of 128\,ns resolution, and the same formula applies:

\begin{equation}
ISN = H_k(4\text{-tuple}) + t \mod 2^{32}
\end{equation}

Accordingly, the same relation $ISN_2 = ISN_1 + (t_2 - t_1) \mod 2^{32}$ applies.

\subsection{Windows}
\label{sec:isn_windows}

Windows presents the most complex case, as its ISN generation algorithm is not standard, and its TCP/IP stack is not open source. To analyze its ISN generation mechanism, we combined black-box measurements with reverse engineering of \texttt{tcpip.sys}, the kernel driver responsible for TCP functionality. In particular, we analyzed the function \texttt{TcpCreateAndConnectTcb\-Rate\-Limit\-Complete}, which generates the ISN. 
Our analysis indicates that the ISN is derived from a keyed hash over the connection's 4-tuple,  combined with an additional offset component. 

We capture this behavior abstractly as: 
\[
\text{ISN} = H_k(4\text{-tuple}) + \text{offset} \mod 2^{32}
\]
Here, $H_k(\cdot)$ denotes a keyed hash function over the connection 4-tuple, where $k$ represents a secret key generated by the kernel at system startup. The additive term (offset) is significantly more complex: while we observe that it includes a time-dependent component, our analysis suggests that it is also influenced by recent connection activity and likely incorporates a random component. 
Importantly for our attack, our reverse engineering indicates that this offset is maintained through four internal ISN-generator buckets, each with its own evolving state.

Building on this observation, we find that the selected bucket is determined by a hash-dependent value derived from the connection 4-tuple. When keeping the source IP, destination IP, and destination port fixed, two source ports that map to the same bucket and are used for connections initiated within a sub-millisecond interval have additive offset components derived from nearly the same bucket state and therefore tend to be close. Since the hash-dependent component is deterministic for each 4-tuple, the difference between the resulting ISNs remains approximately constant across repeated measurements. We define such pairs $(p_1,p_2)$ as \emph{correlative ports}. We also verified this property empirically and found that, for a given $p_1$, approximately $25\%$ of candidate ports are correlative.

While this effect can also occur for $p_1 = p_2$, such cases are not useful for our attack because the malicious app cannot bind to source port $p_1$ in attack time as it is occupied by the victim application, and so they are not considered further. 

\subsection{Exploiting ISN Generation (non-Windows)}
The key observation is 
that for a fixed 4-tuple, the hash component of the ISN remains constant, and only the time-dependent offset changes. Thus, if an attacker obtains the ISN of a connection with a known 4-tuple at a known time, it can infer the ISN of a future connection using the same 4-tuple by accounting for the elapsed time. Accordingly, our goal reduces to obtaining a single reference ISN together with its timestamp. The main challenge is that the ISN is not directly accessible from user space; as we show next, this can be overcome using indirect leakage mechanisms.

\section{TCP Hijacking Attack Details}

In this section, we describe a TCP connection hijacking attack enabled by our approach. By inferring the required connection parameters, an attacker can interfere with a future TCP connection. As a concrete example, we demonstrate an attack targeting  an HTTP connection to port 80 on a remote server (192.0.2.123), in which the attacker injects a forged HTTP response to redirect the victim to attacker-controlled content.

\subsection{Attack Overview}
The attack combines two complementary capabilities: an unprivileged malicious application 
on the victim infers the TCP connection state, while the ARN  injects spoofed packets (in the IP options variant, the ARN is also used for capturing the TCP ISN). As detailed in \autoref{sec:threat_model}, both remain off-path and cannot directly observe the victim--server traffic.
The attack proceeds in four stages, whose details are provided in the subsequent subsections:
\begin{enumerate}
    \item \textbf{Preprocessing.}
  The malicious application constructs a per-port reference table. For each client port $p$, the table stores an ISN and its corresponding timestamp for a connection defined by the 4-tuple (src IP, src port = $p$, dst IP = 192.0.2.123, dst port = 80), where port 80 is used as the standard HTTP service targeted by the attack. We assume the source IP of the victim machine remains fixed for the duration of the attack (i.e., the device does not switch networks). To populate this table, the application repeatedly initiates connections using \texttt{connect()}. For each attempt, it records a timestamp immediately before the call, approximating the moment the SYN packet is generated (and thus the ISN is assigned), and obtains the corresponding ISN via side-channel mechanisms such as cBPF-based leakage or, in the case of IP options, by routing the packets through the ARN, which observes them. 
  The table is constructed for all candidate client ports, since the source port used by the victim's connection is not known in advance. This table is later used to infer the ISN value of a  connection attempt to 192.0.2.123:80. 

    \item \textbf{Port discovery.} 
    The malicious code detects when a new connection to the target server is initiated and identifies the allocated client port, together with a timestamp close to connection initiation. In our setting, this is achieved primarily using two techniques: monitoring operating-system connection state via interfaces such as \texttt{/proc/net} to directly identify allocated ports, and exploiting ephemeral port allocation patterns to predict the port used by newly established connections.
    \item \textbf{ISN inference.} Using the discovered client port and its associated timestamp, the malicious code consults the precomputed reference table to retrieve the corresponding ISN sample and adjusts it based on the elapsed time to infer an approximation of the target connection's ISN. Since the exact ISN generation time cannot be determined, this process yields a range of candidate values around the genuine ISN. These candidates are then transmitted to the remote adversary.
    \item \textbf{Packet injection.} Using the inferred ISN candidates and the known 4-tuple, the remote adversary sends spoofed packets that are accepted by the victim as part of the legitimate connection, thereby hijacking it.
\end{enumerate}

\subsection{Pre-Processing}
We consider two independent methods for constructing the port--ISN--timestamp table: one based on cBPF (applicable to IPv4 and IPv6 on Linux and Android) and one based on IP options (IPv4, cross-platform).

\subsubsection{cBPF Pre-Processing}
 In our setting, the malicious application uses cBPF as a leakage primitive to recover the client's ISN from packets belonging to an established connection to the target server. For each connection, it attaches a cBPF filter that inspects the TCP header of an incoming packet, and, by reading from a fixed offset, obtains a selected byte of the acknowledgment number and encodes it into the returned packet length. As a result, the number of bytes delivered to user space directly reveals the value of that byte. 

To reconstruct the full 32-bit ISN, the malicious application repeats this process four times over the same connection. 
In each iteration, the malicious application issues an HTTP request and attaches a cBPF filter configured to leak a different byte of the acknowledgment number from the corresponding response packet.
The acknowledgment number reflects the client's sequence number, i.e., the client's ISN plus an offset corresponding to the data already sent. By leaking the acknowledgment number byte after byte, the malicious application reconstructs the original ISN. For each sampled connection, the application also records the timestamp of the initial SYN packet, i.e., the time at which the ISN is generated. This process is repeated for all candidate client ports.\footnote{
\label{fn:reduce_preprocessing_time} To reduce preprocessing time, we restrict probing to the default
Linux ephemeral-port range, \texttt{32768--60999}, and consider only ports
with the parity used by the TCP ephemeral-port allocator. For this default
range, \texttt{connect()} selects even parity source ports because the
lower bound is even. This behavior follows from the Linux source port
allocation logic in \texttt{\_\_inet\_hash\_connect()}~\cite{linux-inet-hashtables}.}

\subsubsection{IP Options Pre-Processing}
\label{sec:ip-option-preprocessing}
In this approach, we use IPv4 source routing to force connection initiation packets whose final destination is the service's IP address and port, to traverse an attacker-controlled machine (ARN). For each source port $p$, the malicious application initiates a TCP connection to the target server using a 
\texttt{connect()} call, which triggers the transmission of a SYN packet. 
\ifextended
We use a non-blocking call to trigger SYN transmission without waiting for connection completion, thereby saving time during the preprocessing phase. The socket is configured with the LSRR (Loose Source Routing) option using \texttt{setsockopt(\ldots, IPPROTO\_IP, IP\_OPTIONS, \ldots)}, causing the SYN packet to be routed through the ARN. 
\fi
By capturing this SYN packet, the ARN obtains the client-side sequence number and sends it to the malicious application. The corresponding timestamp is recorded by the malicious application immediately before invoking \texttt{connect()}, approximating the time at which the SYN packet is generated and the ISN is assigned. This process is repeated for all candidate source ports. As a result, the malicious application constructs a reference table that maps each source port to its corresponding ISN and timestamp.

In practice, many routers filter or ignore IP options, limiting the applicability of this technique. However, a portion of network paths still permit such packets \cite{ip_options_supported}.
Therefore, we demonstrate a weaker setting in which the ARN resides in the local switched network as the victim (i.e. no router between it and the victim machine), while remaining off-path, and has elevated privileges that allow it to inspect entire packets destined to it (including e.g. packets explicitly routed through it via IP options). Notably, our approach does not rely on traffic interception techniques such as ARP spoofing or similar methods that redirect all victim traffic through the attacker.

\subsection{Source Port Discovery}
\label{sec:source_port_discovery}
Our goal is twofold: (i) to identify the source port used by a newly established connection, and (ii) to obtain a timestamp as close as possible to the moment the connection is initiated (i.e., when the ISN is generated). These two pieces of information are required in order to later infer the ISN as accurately as possible. Below we list the techniques we developed for port inference, and in \autoref{tab:os-methods} we depict the preferred method for each operating system (in terms of response time). 
\subsubsection{Method 1: OS-Level Connection Monitoring}
\label{sec:os-level-connection-monitoring}
The first approach relies on querying operating system APIs to monitor active connections. On Linux, this can be done by repeatedly reading \texttt{/proc/net/tcp}, part of the procfs pseudo-filesystem mounted under \texttt{/proc}  that exposes kernel information (e.g., networking and process state) to user space. On Windows, similar information can be obtained via APIs such as \texttt{GetExtended\-Tcp\-Table}. By parsing these tables, one can detect newly established connections and identify the allocated source port.

However, parsing the full connection table incurs a significant overhead and limits timing resolution. To improve efficiency on Linux, we instead monitor the \texttt{ActiveOpens} counter (pseudo-file) in \texttt{/proc/net/snmp}, which tracks the number of outgoing TCP connections initiated by the host (regardless of their destination). Since this pseudo-file has a fixed small size, it can be polled at high frequency with minimal overhead, enabling more precise detection of connection initiation events.\footnote{We carefully optimized the detection of connection initiation timing, as even small inaccuracies can significantly affect the predicted ISN. In particular, due to the timer granularity of 64ns, 
an error of 1ms in estimating the connection initiation time may result in an ISN prediction error of approximately 16{,}000 in Linux.} Once a new connection is detected and its approximated timestamp is recorded, the corresponding port can be identified via procfs. 
This method is not applicable on Android, where regular 
applications do not have access to \texttt{/proc/net} entries.

\subsubsection{Method 2: Exploiting Linux TCP Ephemeral Port Selection}
\label{sec:linux-ephemeral-port-selection}

To overcome the limitations in Android mentioned above, 
we exploit the Linux TCP ephemeral port selection algorithm, which is an adaptation of Algorithm 4 in RFC 6056 \cite{rfc6056}. Linux selects ephemeral TCP source ports from a configurable range, which by default is 32768--60999, 
and uses a perturbation table \texttt{table\_perturb[]} of fixed size (65536 entries), indexed by a hash of the connection 3-tuple, defined as $(\text{src IP}, \text{dst IP}, \text{dst port})$. For a given 3-tuple $\tau$, 
the kernel computes: 
\[
base\_port = \bigl(Table[h_1(\tau)] + h_2(\tau)\bigr) \bmod R + min\_port
\]
Here, $R$ denotes the size of the ephemeral port range, and $min\_port$ denotes its lower bound. As noted above, on Linux, this range is configurable and is typically 32768--60999 by default, yielding $R = 28232$. If $base\_port$ is occupied, the kernel probes $base\_port+2, base\_port+4, base\_port+6, \ldots$ until an available port is found. After a port is allocated, the table entry is updated as follows (this is a simplified logic assuming $base\_port$ was available):

\[
Table[h_1(\tau)] \mathrel{+}= 2 \cdot r
\]
(where $r \in \{1,\ldots,8\}$ is random).\\
As a result, for a fixed 3-tuple, consecutive connections are assigned ports that differ by one of a small set of values. In particular, if a connection is assigned port $p$, the next port will be one of:
\[
\{p + 2, p + 4, p+6, \dots, p + 16\}.
\]We leverage this structure as follows. After obtaining a port $p$ via a \texttt{connect()} call to the target server (thus using the same 3-tuple as the victim, as we assume the source IP does not change during the attack), we proactively occupy the ports $\{p+2, p+4, \dots, p+14\}$ using \texttt{bind()} invocations. As a result, the next available source port candidate for the 3-tuple is $p+16$, which the kernel is forced to select. We then attempt to \texttt{bind()} to port $p+16$; if the call fails, it indicates that the port has just been allocated by the victim, allowing us to infer both the source port ($p+16$), and an approximate timestamp of the connection establishment.

A practical complication is that the cryptographic keys used in the hash functions are rotated periodically  
(every 10 seconds). This causes abrupt changes in the observed port sequence, since once it changes, we are no longer sampling the correct port that will be assigned. To handle this, we continuously 
establish connections to a reference host with fixed destination port (e.g. port 80) 
and monitor the source port sequence: as long as the key is unaltered, source ports increase predictably; a sudden deviation indicates re-keying, at which point we resynchronize our measurements. 

\subsubsection{Method 3: Exploiting macOS, iOS, and Windows TCP Ephemeral Port Selection}
\label{sec:cross-platform-ephemeral-port-selection}

On XNU-based operating systems (macOS and iOS), ephemeral TCP ports increment sequentially. This behavior is reflected in the XNU kernel's port allocation logic (function \texttt{in\_pcbbind()}~\cite{xnu_inpcb}). On Windows, our empirical evaluation indicates that ephemeral TCP ports increment sequentially over short time intervals, after which a new random base port is selected and the sequence continues. Despite these resets, this behavior remains exploitable, as the base port can be probed frequently to track the sequence.

This significantly simplifies port prediction, as the next port can be inferred with high confidence from a single sample. 

\subsubsection{Method 4: bind()-based probing}
\label{sec:bind-port-scanning}
A straightforward approach is to attempt binding to all possible ports. We do this sequentially, attempting to bind to each port and quickly releasing it in order not to exhaust the entire port space. 
We refer to this systematic enumeration approach as bind-all probing. 
If a \texttt{bind()} call fails, the corresponding port is already in use. While simple, this method is inefficient and incurs significant overhead. As shown in \autoref{tab:port-scan-time}, due to its low speed, we use this method only for the DNS cache poisoning attack and not for TCP hijacking.

\subsubsection{Method 5: Ephemeral Port Sampling via bind(0)}
\label{sec:bind-zero-port-sampling}
Finally, we consider invoking \texttt{bind()} with a zero port (herein \texttt{bind(0)}), which instructs the kernel to assign an available ephemeral port to the socket. By repeatedly invoking this operation and immediately releasing the port, we obtain samples from the set of currently available ports. Ports that are never assigned during this process are likely to be in use. This is a probabilistic technique which can result in false positives (ports that are never assigned to us). 
However, as shown in \refbindzeroanalysis, with sufficiently many samples, the probability of missing available ports becomes negligible.

Consequently, the sampled set is typically sufficient to detect the source port in practice. This approach is clearly inferior to the \texttt{bind()}-based probing method, which systematically scans all ports and thus provides complete and efficient coverage. However, our motivation in exploring this technique is to understand the attacker's capabilities in more restricted environments, where (hypothetically) interfaces such as procfs are unavailable and binding to user-designated ports is not permitted. In such settings, \texttt{bind(0)} remains one of the few available primitives, making this approach relevant despite its limitations. Because of its low speed, as shown in \autoref{tab:port-scan-time}, we use this method only for the DNS cache poisoning attack and not for TCP hijacking.

\begin{table}[t]
\centering
\caption{Time per occupied-port detection iteration on Linux}
\label{tab:port-scan-time}
\begin{minipage}{\columnwidth}
\centering
\begin{tabular}{l l l}
\toprule
\textbf{Method} & \shortstack{\textbf{Average Time}\\\textbf{per Iteration}}
& \shortstack{\textbf{Std. Dev.}}\\
\midrule
1 --- procfs monitoring & $0.04\,\mathrm{ms}$ & $0.006\,\mathrm{ms}$ \\
\begin{tabular}[c]{@{}l@{}} 2 --- TCP Ephemeral port\\ selection exploitation\end{tabular} & $0.006\,\mathrm{ms}$\footnote{In practice, when another process simultaneously invokes \texttt{bind()}, there may be a small spike in reaction time here, as explained in \reftimesidechannel. 
} & $0.0087\,\mathrm{ms}$\\
4 --- \texttt{bind()}-based probing\footnote{For the \texttt{bind()} and \texttt{bind(0)} probing methods, we scanned only ports 32768--60999, corresponding to the default Linux 
ephemeral UDP-port range. We use this range because these methods are later applied to identify the UDP source port used in the DNS cache-poisoning attack.}\ & $46\,\mathrm{ms}$ & $1\,\mathrm{ms}$\\
5 --- \texttt{bind(0)} sampling & $429\,\mathrm{ms}$ & $223\,\mathrm{ms}$\\
\bottomrule
\end{tabular}
\end{minipage}
\end{table}

\begin{table}[t]
\centering
\caption{Preferred source port discovery method per operating system}
\label{tab:os-methods}
\begin{minipage}{\linewidth}
\begin{tabular}{l l l l}
\toprule
\textbf{OS} & \textbf{Preferred Method} & \shortstack{\textbf{Avg.}\\ \textbf{Time}} & \shortstack{\textbf{Std.}\\ \shortstack{\textbf{Dev.}}} \\
\midrule
Linux & procfs monitoring &  $40\,\mu s$  & $6\,\mu s$ \\
Android & \begin{tabular}[c]{@{}l@{}}TCP Ephemeral port\\ selection exploitation\end{tabular}   &$153\,\mu s$  & $70\,\mu s$ \\
Android & \texttt{bind()}-based probing\footnote{We scanned only ports 1025--65534, corresponding to the default Android ephemeral UDP port range for DNS, because this method is used to identify the UDP source port in the DNS cache poisoning attack.} & $179\,\mathrm{ms}$ & $14\,\mathrm{ms}$ \\
macOS & Sequential TCP port prediction & $276\,\mu s$  & $334\,\mu s$ \\
iOS & Sequential TCP port prediction & $222\,\mu s$  & $145\,\mu s$\\
Windows & OS-level table monitoring & $72\,\mu s$ & $9.6\,\mu s$  \\
\bottomrule
\end{tabular}
\end{minipage}
\end{table}

\subsection{TCP Hijacking (non-Windows)}
\label{sec:tcp-hijacking}

The TCP hijacking phase is performed by actively interfering with the TCP handshake, impersonating the legitimate server and sending forged SYN-ACK packets to the victim. We target this stage because, after the handshake completes, hijacking becomes significantly more difficult: while the attacker may infer the client's sequence number, the server's sequence number remains unknown.

For a forged SYN-ACK packet to be accepted, the acknowledgment number must match the expected value, i.e., $\text{ack\_num}_{\text{SYN-ACK}} = \text{seq\_num}_{\text{SYN}}+1 = \text{ISN} + 1$. 
Since our prediction granularity of ISN is a range of typically hundreds to thousands values, depending on the operating system, the attacker needs to perform a brute-force search over this range by sending multiple candidate SYN-ACK packets. 

After the victim initiates a connection (detected via the occupied port), the attacker waits briefly, on the order of a few milliseconds, to allow the legitimate SYN to be transmitted toward the genuine server. The attacker then injects spoofed SYN-ACK packets. Specifically, the attacker performs a brute-force search over the TCP acknowledgment number field, iterating over candidate values derived from the inferred client ISN. For each guess, a corresponding SYN-ACK packet is sent. The sequence number in these packets can be set to an arbitrary fixed value, as the attacker controls the forged server side of the connection. During this phase, the victim responds with RST packets to incorrect guesses, until a SYN-ACK with the correct acknowledgment number is received. 

Once the victim receives a valid SYN-ACK (the attacker's one, if it wins the race against the genuine server's), the victim replies with a TCP ACK packet, completing the handshake, after which the client issues its HTTP request. The attacker then performs a second brute-force phase using forged PSH-ACK packets carrying a malicious HTTP response. Unlike the SYN-ACK phase, this stage does not require an exact match of the acknowledgment number. Instead, it suffices for the acknowledgment number to fall within a valid acceptance range defined in the TCP protocol.\footnote{We use increments equal to (or smaller than) the size of the victim application's last HTTP request, i.e., $\text{SND.NXT} - \text{SND.UNA}$, which defines the window of valid acknowledgment numbers.}
Therefore, the attacker can test candidate acknowledgment numbers using larger step sizes, reducing the number of required guesses. Once a valid TCP packet is accepted, the victim application processes the injected HTTP response. 

For the attack to succeed, the 
SYN-ACK packet injected by the attacker with the correct acknowledgment number must arrive at the victim before the legitimate SYN-ACK sent by the genuine server. 

Thus, the attack succeeds when the following condition is satisfied:
\[
\text{RTT}_{\text{victim}\leftrightarrow\text{ARN}} + T_{\text{brute\_force}} < \text{RTT}_{\text{victim}\leftrightarrow\text{server}}.
\]

\ifextended
Furthermore, once the client's socket transitions into the ESTABLISHED state, each ACK it sends (both the one that concludes the 3-way handshake, and those sent in response to incoming data) will cause the server (in SYN\_RECEIVED state) to send back an RST packet whose sequence number is copied from the ACK's acknowledgment number (which is the client's RCV.NXT at the time the ACK was sent). If no further data was received at the client during the time between the ACK was sent and the RST was received, then RCV.NXT is unmodified, and the RST's sequence number will match it, which will result in the client accepting the RST and terminating the connection. Conversely, as long as data (payload) is accepted by the client no longer than $\text{RTT}_{\text{victim}\leftrightarrow\text{server}}$ time after the last ACK was sent, the RST packet's sequence number will be \emph{lower} than the client's RCV.NXT, and therefore the RST will be silently discarded and the connection will remain open. The attacker can gracefully terminate the connection by sending FIN-ACK 
(which increments the client's RCV.NXT by one, thus again making the server's in-flight RSTs ineffective). 
\fi 

A detailed diagram of the attack flow is provided in \reftcphijackingflow.

This condition also clarifies the effect of low-latency and CDN-fronted servers. A nearby CDN edge may reduce $\mathrm{RTT}_{\mathrm{victim}\leftrightarrow\mathrm{server}}$, thereby narrowing the attacker's race window and, in sufficiently low-latency settings, making the attack impractical. However, CDN deployment does not inherently imply a short TCP-handshake latency. 
In an additional measurement of an HTTP connection 
from a home connection in \anonymize{REDACTED}{Israel} 
to a Verizon service fronted by Akamai, approximately $170\,\mathrm{ms}$ elapsed between transmission of the SYN and receipt of the SYN-ACK. 
This delay leaves a substantial race window for an ARN located close to the victim, showing that CDN fronting does not necessarily prevent the attack.

\subsection{TCP Hijacking on Windows}
\label{sec:isn_attack_windows}

Based on the observed properties of the ISN generation process described in \autoref{sec:isn_windows}, we design an attack that leverages correlations between selected port pairs. 
\begin{enumerate}
    \item \textbf{Preprocessing.}
    For each source port $p_1$, the malicious application invokes \texttt{bind()} to bind to $p_1$, and then calls \texttt{connect()} to initiate a TCP connection to the target service IP and port. It then rapidly initiates another connection using a candidate source port $p_2$, selected from a small set of base ports, with a sub-millisecond delay between the two attempts. The remaining elements of the 4-tuple are kept fixed. The SYN packets include IP options so that they can be captured by the ARN. The ARN extracts the two ISNs, denoted $\mathrm{ISN}(p_1)$ and $\mathrm{ISN}(p_2)$, and computes their difference
    \[
    \Delta_{p_1, p_2} = \mathrm{ISN}(p_1) - \mathrm{ISN}(p_2) \pmod{2^{32}}.
    \]
    This process is repeated for many candidate $p_2$ values and across multiple trials. If the difference $\Delta_{p_1, p_2}$ remains approximately constant across trials, the pair $(p_1, p_2)$ is marked as correlative. The preprocessing phase therefore produces a reference table of triples $(p_1,p_2,\Delta_{p_1, p_2})$, where $\Delta_{p_1, p_2}$ is the stable ISN difference associated with that port pair. In practice, many such correlative pairs exist for each $p_1$ (recall that roughly 25\% of the pairs are correlative).

    \item \textbf{Attack.}
    The victim application initiates a connection using port $p_1$ (and with ISN=$\text{ISN}_1$). 
    As soon as a SYN packet is sent (as part of the TCP handshake), the malicious application detects the source port $p_1$ (via the Windows equivalent of \texttt{/proc}-based monitoring, as described in \autoref{sec:os-level-connection-monitoring}) 
    and immediately initiates its own TCP connection from the correlative source port $p_2$ using a \texttt{connect()} call.
     
    The SYN packet of the malicious application (with $\text{ISN}_2$) is sent with IP options enabled so that it can be captured by the ARN. The ARN extracts $\text{ISN}_2$ and, using the inferred source port $p_1$ and the precomputed offset $\Delta_{p_1, p_2}$ (i.e., the approximately constant ISN difference associated with the port pair $(p_1,p_2)$), estimates the victim's ISN as $\text{ISN}_1 \approx \text{ISN}_2 - \Delta_{p_1, p_2} \mod 2^{32}$.

    The remainder of the attack proceeds as in the above subsections.
\end{enumerate}

\section{DNS Cache Poisoning Attack details}
We demonstrate a DNS cache poisoning attack against the target device's stub resolver, using our source UDP port discovery techniques. In this setting, we exploit the fact that a forged DNS response is accepted if it matches the query's validation fields, most importantly the UDP source port and the DNS transaction ID (TXID). All remaining fields---such as the queried domain name, query type, and the contents of the answer, authority, and additional sections---are either known in advance or can be freely chosen by the attacker.

\subsection{Attacking Linux}
\label{sec:attacking-linux}
Linux systems widely deploy systemd-resolved as the local stub resolver. In particular, it is the default resolver in Ubuntu---the most popular Linux distribution \cite{ubuntu-popular}---as well as in other major distributions such as Fedora and RHEL. In our experiments, we demonstrated the attack on Ubuntu using systemd-resolved as the local stub resolver. This resolver was particularly convenient for demonstrating the attack because it follows a record-based caching model \cite{klein_ndss_2019},  
whereby individual resource records contained in a DNS response are cached separately, regardless of the original query name. 
Notably, a recursive resolver response may legitimately contain records regarding domains other than the one queried, as is the case with CNAME chains. Such answers are by-definition trusted and cached by the stub resolver. 
As a result, an attacker can construct responses that appear fully valid yet introduce malicious records (irrespective of the original query domain), which may be cached and later served, enabling cache poisoning.

In that, our attack approach follows the principles of prior DNS cache poisoning attacks against stub resolvers, described in the literature  \cite{cross_layers_attack_klein,
weak_randomness_android}.

Using the previously described techniques for detecting allocated ports on the victim (e.g., monitoring \texttt{/proc}, exhaustive bind probing, and \texttt{bind(0)} sampling), we identify the UDP source port used by the stub resolver when communicating with the recursive resolver.

Notably, Method~2 (exploiting the ephemeral port selection algorithm, described in \autoref{sec:linux-ephemeral-port-selection}) is not applicable in this context, as the Linux UDP source port selection method does not make use of the algorithm described for TCP earlier and instead selects ports at random.

Suppose the attacker aims to poison the cached record for \texttt{example.com}. In the first step of the attack, the malicious application running on the victim triggers a DNS query for a domain under attacker control (e.g., \texttt{random.\allowbreak attacker.\allowbreak test}), 
infers the stub resolver's UDP source port for this query and transmits it to the ARN. 
The ARN then sends spoofed replies, impersonating the recursive resolver (i.e., using the latter's source IP address), with different candidate TXIDs in an attempt to match the query. Since the query is issued to an attacker-controlled domain, the attacker can delay the legitimate response, effectively extending the time window available for the brute-force TXID search.
The injected response includes records such as:
\begin{verbatim}
random.attacker.test   CNAME   example.com
example.com            A       192.0.2.66
\end{verbatim}
As a result, if one of the forged replies contains the correct TXID and arrives before the genuine response (which is delayed), the stub resolver accepts it and caches a poisoned mapping for \texttt{example.com}, 
(overriding an existing genuine cached record if it exists) causing subsequent connections to \texttt{example.\allowbreak com}  
to be redirected to an attacker-controlled address.

A practical limitation arises with the bind-all 
and \texttt{bind(0)}-based methods, which, as shown in \autoref{tab:port-scan-time}, are relatively slow and take more than 40ms per iteration. When the malicious application initiates a DNS query, two ports become occupied: one ephemeral port used for loopback communication to the systemd-resolved stub resolver, and another used by systemd-resolved to communicate with the recursive resolver. The latter is the relevant port for the attack; however, due to the coarse sampling interval, we cannot reliably distinguish between the two.
Consequently, we must attack both ports (brute-force the TXID space in each of them). This increases the time of the attack and lowers its  success rate per round. In order to still ensure we poison the cache, the attack can be repeated multiple times in succession. In contrast, when using procfs, this complication does not arise, as the stub resolver's loopback source port is detected first, then the outbound port is detected (separately), and we can thus distinguish among them and target the latter one. So, considering the procfs variant, the attack succeeds in a single round, provided that $\text{RTT}_{\text{victim}\leftrightarrow\text{ARN}}+65536 \cdot R < \text{Timeout}$, where $R$ is the ARN spoofed DNS answer transmission rate, typically hundreds of thousands (packets) per second, and $\text{Timeout}$ is the recursive resolver timeout (the minimal recursive resolver timeout is 1 second \cite[Table II]{cross_layers_attack_klein}). Therefore, this condition is easily met in real world scenarios.

\subsection{Attacking Android}
\label{subsec:android-dns}

Android employs a query-based DNS cache rather than the record-based
cache used by 
systemd-resolved. Consequently, the Linux
technique described in the previous subsection, in which the attacker
injects additional records through a forged CNAME chain, is not
applicable to Android. 
A DNS cache poisoning attack against Android begins, as in the Linux
case, by triggering a DNS query and identifying the UDP source port
used by the local stub resolver for that query (randomly selected from
the range 1025--65534~\cite{aosp_dns_random_port}). Once the source port is
known, the ARN races the genuine recursive resolver by
sending forged responses containing different candidate TXIDs.

Because Android caches responses according to the queried domain, the
malicious application must trigger a query for the actual domain that
the attacker wishes to poison, for example,
\texttt{example.\allowbreak com}. This setting is more challenging than
the systemd-resolved case. In the Linux attack, the query is
issued for an attacker-controlled domain, allowing the attacker to delay
the legitimate response and thereby extend the time available for
source-port identification and TXID brute forcing. On Android, the
attacker cannot delay the genuine response and must complete both stages
within the regular DNS response time.

Moreover, each query provides only a single attack opportunity. If none
of the forged responses contains the correct TXID before the legitimate
response arrives, the genuine record is inserted into the cache. As long
as this record remains cached, subsequent requests are answered locally
and do not generate a new upstream DNS query.

An unprivileged Android application cannot directly flush the DNS cache,
because the relevant cache-management operations require elevated
privileges. Nevertheless, the attacker can attempt to evict the target
entry by issuing queries for a sufficiently large number 
 of other 
domains (the default cache capacity is 640 entries~\cite{aosp_dns_cache}). After the target entry is evicted, the malicious application can trigger another query for the target domain and repeat the attack. 

A further complication concerns source-port discovery. Unlike on Linux, an unprivileged Android application cannot inspect
\texttt{/proc/net}. Therefore, among the techniques considered in this work, the attacker
must rely on \texttt{bind()}-based probing (or the slower \texttt{bind(0)}-based probing). 
A full \texttt{bind()}-based scan on Android takes approximately $179\,\mathrm{ms}$ (as shown in \autoref{tab:os-methods}), which significantly exceeds the time required for the genuine DNS response to arrive (typically 50--100~ms).
The attacker must therefore scan only a
subset of the candidate ports and use the remaining time to brute-force
the TXID (only a portion of the TXID space, due to time constraints). Consequently, an individual attack attempt may fail for either
of two reasons: the scanned subset may not include the stub resolver's source
port, or the transmitted forged responses may not include the correct
TXID.

Let $T=\text{RTT}_{\text{victim}\leftrightarrow\text{resolver}}-\text{RTT}_{\text{victim}\leftrightarrow\text{ARN}}$ denote the effective attack window. The attacker allocates $x$
time units to source-port probing and $T-x$ time units to TXID brute
forcing. Under the simplifying assumption that both the UDP source port
and the TXID are uniformly distributed, the approximate success
probability of a single attempt is
\begin{equation}
    P_{\mathrm{succ}}(x)
    =
    \frac{A \cdot x}{65534-1025+1}
    \cdot
    \frac{R \cdot (T-x)}{65536},
    \label{eq:android-dns-success}
\end{equation}
where $A$ is the port-probing rate (ports/sec) and $R$ is the forged-packet
transmission rate (packets/sec). 
Since all factors other than $x(T-x)$ are constant, maximizing
\autoref{eq:android-dns-success} is equivalent to maximizing $ x(T-x)$, which attains its maximum at $ x = \nicefrac{T}{2}.$

Thus, the attacker should divide the
available race window equally between source-port probing and TXID
brute forcing. In a single attack attempt, therefore, the success probability is 
\begin{equation}
    P_{\mathrm{succ}}
    =
    \frac{A \cdot R \cdot (\text{RTT}_{\text{victim}\leftrightarrow\text{resolver}}-\text{RTT}_{\text{victim}\leftrightarrow\text{ARN}})^2}{4 \cdot (65534-1025+1)\cdot 65536}
\end{equation}

Although a single attempt has limited success probability, the attacker can evict the cached entry and repeat the attack until success. Note that each individual attempt takes $\text{RTT}_{\text{victim}\leftrightarrow\text{resolver}}+\text{T}_{\text{evict}}$ time.

\subsection{Attacking Windows}

The Windows attack is more similar to the Android attack than to the
Linux systemd-resolved attack. Windows employs a query-based
DNS cache, and therefore the attacker must trigger a query for the
actual target domain and complete the attack within the short interval
before the genuine response arrives.

Unlike Android, an unprivileged Windows application can immediately
identify the resolver's UDP source port (randomly selected from the default port range
49152--65535~\cite{microsoft_dns_ports}) using
\texttt{GetExtendedUdpTable}, as described in
\autoref{sec:os-level-connection-monitoring}. Cache eviction is also
simpler: the attacker can flush the entire DNS cache using the
undocumented \texttt{dnsapi.DnsFlushResolverCache} function, which is
available to unprivileged applications. 

Windows introduces an additional complication. Prior
work~\cite[Section~V]{collaborative_client_side_dns} shows that the
Windows stub resolver terminates the outstanding query after
receiving five responses with incorrect TXIDs (but with correct ports). We additionally observed
that even a single response with an incorrect TXID prevents a subsequent valid answer from being cached, although it does not terminate the
outstanding query. Since the attacker requires the forged response to
enter the cache, only one TXID candidate can be tested in each attack
attempt.
For a single attempt, we only require $\text{RTT}_{\text{victim}\leftrightarrow\text{ARN}} < \text{RTT}_{\text{victim}\leftrightarrow\text{resolver}}$, as port inference time is negligible, and so is the time to transmit 5 spoofed answers (by the ARN). The probability of a single attempt to succeed is $\nicefrac{1}{65536}$ (the probability of the guessed TXID to be correct).

If the attempt fails (i.e. the attacker's TXID guess is incorrect), Windows returns an error code (\texttt{DNS\_ERROR\_BAD\_PACKET}) and the attacker can immediately start a new attempt, until success. Each attack round takes $\text{RTT}_{\text{victim}\leftrightarrow\text{ARN}}$. Note that since in Windows the attacker terminates every query before the genuine response arrives from the resolver, the attacker only needs to flush the DNS cache once, before the attack begins, in order to remove a cached entry (if any) for the domain under attack. In our code, however, we flushed the cache after each attack attempt failure, just in case (the flushing time in Windows is negligible: $\approx 140\mu s$).

\subsection{Attacking macOS and iOS }
macOS and iOS use \texttt{mDNSResponder} as the system stub resolver. In principle, the same initial step of identifying the UDP source port would apply. However, 
\texttt{mDNSResponder} enters a defensive mode after receiving even a single invalid DNS response for an outstanding query (for example, a reply with an incorrect TXID). Once triggered, the resolver retries subsequent lookups over TCP rather than UDP for approximately 10 seconds. Because TCP responses cannot be spoofed using the same off-path brute-force method, this behavior effectively prevents the attack.

\section{Experiments}

Our experiments are logically divided into three groups:

\begin{enumerate}
    
    \item \textbf{TCP hijacking using cBPF pre-processing.} cBPF is available only on Linux and Android. 
   \item \textbf{TCP hijacking using IP-options pre-processing.} Here, we demonstrated the attack on Linux, Android, macOS, iOS, and Windows. Because our routers and upstream ISP infrastructure do not reliably support IP options, we split the attacker remote node (ARN) into two components: (i) a \emph{capturing-ARN}, placed within the local network, responsible for receiving and inspecting packets explicitly routed through it via IP options sent by the victim; and (ii) an \emph{injecting-ARN}, responsible for sending spoofed packets. For Linux and Android, the injecting-ARN was hosted on AWS, demonstrating that remote injection is feasible once the capturing-ARN on the local network receives the victim's ISN. For macOS, iOS, and Windows, the  injecting-ARN was placed within the local network in order to simplify the experimental setup.

    \item \textbf{DNS cache poisoning.} We demonstrated DNS cache poisoning attacks on Linux, Android, and Windows. On Linux, we evaluated three source-port discovery methods: \texttt{/proc}-based monitoring, binding all ports, and \texttt{bind(0)}.
    
\end{enumerate}

Details of the client platforms used in the experiments are provided in \autoref{tab:platforms}. Our success ratios for the TCP hijacking attack are presented in \autoref{tab:tcp_hijack_results}. 
We hypothesize that most failed attack attempts result from 
an error in ISN inference, causing the true ISN to fall outside the candidate range tested by the ARN.
In all experiments, communication between the malicious application and the ARN was implemented using simple UDP messages.

\begin{table}[h]
\setlength{\tabcolsep}{1pt}
\centering
\caption{Tested platforms and configurations}
\label{tab:platforms}
\begin{tabular}{ c c c }
\hline
OS & Version & Device \\
\hline
Linux & 7.0.0-rc6 (Ubuntu 24.04.4) & Dell mini PC (i7-13700T) \\
Linux & 7.0.0-rc1 (Ubuntu 24.04 LTS) & Azure VM (B2ts v2) \\
Android & Android 16 (Kernel 6.1.145) & Google Pixel 9 \\
Windows & Windows 11 Home, ver. 25H2 & Dell mini PC (i7-13700T) \\
macOS & macOS Tahoe 26.4 & MacBook Air (M2) \\
iOS & iOS 26.3.1 & iPhone 15 \\
\hline
\end{tabular}
\end{table}

 \subsection{Local Network Setup for IP Address Spoofing}
In our experimental setup, the attacker remote node (ARN) was hosted on an AWS instance, which prevents sending packets with arbitrary (spoofed) source IP addresses. Therefore, the ARN transmitted packets using its own genuine IP address rather than spoofing the address of the target server. To overcome this limitation, we configured the victim's local network accordingly. In the IPv4 setting, the victim was connected to the Internet through a router with a 
port-preserving NAT. On this router, we deployed an XDP/tc filter that rewrites only the source IP address field of incoming packets received from the ARN to match that of the genuine server, without modifying any other packet fields (except for the checksums which are adjusted automatically) or behavior. As a result, packets crafted by the ARN appear to the victim client as if they were sent by the genuine server and are therefore accepted as part of the legitimate flow. In the IPv6 setting, we used a standard router, and the source IP address was rewritten directly on the victim machine. This adjustment was introduced solely due to deployment constraints and does not affect the core attack logic, which remains applicable in any setting where source address spoofing is permitted.

\subsection{Setup for TCP Hijacking --- Linux}
As mentioned earlier, both cBPF pre-processing and IP-options pre-processing were demonstrated.
The cBPF preprocessing took 305 seconds and generated approximately 290 MB of network traffic, whereas the IP-options preprocessing took 22 seconds and generated approximately 2 MB. 

We scanned only the relevant range of even ports, as explained in \autoref{fn:reduce_preprocessing_time}. Source port detection was performed using the \texttt{/proc}-based technique described in \autoref{sec:os-level-connection-monitoring}. We were able to reduce the ISN uncertainty down to 1.6K candidate values which was sufficient to carry out the attack with a high success rate, as shown in \autoref{tab:tcp_hijack_results}. The ARN 
was hosted on an AWS machine. In the IP-options setting, packet capturing was performed by the capturing-ARN in the local network, while the injecting-ARN was an external attacker (AWS).

The victim was a Linux machine located in our lab in \anonymize{REDACTED}{Israel}. The victim application was a simple Python client we implemented using standard sockets to perform a regular HTTP request; although we also demonstrated hijacking browser-based connections, we relied on this client for evaluation to avoid browser caching effects. The network latencies were $\text{RTT}_{\text{victim}\leftrightarrow\text{server}} \approx 108\,\mathrm{ms}$ and $\text{RTT}_{\text{victim}\leftrightarrow\text{ARN}} \approx 5\,\mathrm{ms}$. The injecting ARN transmits 1.6K SYN-ACK packets in 11ms (approximately 145K packets/s), 
 and the bandwidth between the capturing ARN (on the AWS EC2 machine) and the victim was just above 150 Mbit/s (in both directions).

\subsection{Setup for TCP Hijacking --- Android}
Our code (both the pre-processing logic and the malicious application) was implemented as a regular Android application. Both IP-options and cBPF pre-processing techniques were demonstrated. 
Source port detection was performed using the Linux ephemeral port selection method described in 
\autoref{sec:linux-ephemeral-port-selection}.
TCP hijacking --- particularly HTTP response injection --- 
was successfully demonstrated. The cBPF setting was demonstrated with the ARN hosted on an AWS machine ($\text{RTT}_{\text{victim}\leftrightarrow\text{ARN}}$ was $14\,\mathrm{ms}$). In the IP-options setting, the capturing-ARN was placed 
in the local network, while the injecting-ARN was on an AWS machine.
As in the Linux scenario, to avoid browser caching effects, we implemented a simple C++ 
client application using standard sockets.

\subsection{Setup for TCP Hijacking --- macOS and iOS}

The on-device logic was implemented as an unprivileged application on both macOS and iOS. IP-options pre-processing was used. Source port detection was performed by exploiting the platform-specific port selection algorithm, as described in \autoref{sec:cross-platform-ephemeral-port-selection}. The ARN, comprising both the capturing-ARN and the injecting-ARN components, was placed within the local network. For the TCP client, we implemented a simple application using standard sockets (a Python script on macOS and a C++ application on iOS).

\subsection{Setup for TCP Hijacking --- Windows}
We note that Windows may employ additional protections such as Smart App Control (SAC), a protection mechanism backed by Microsoft intelligence services that restricts the execution and behavior of untrusted applications. In our experiments, SAC blocked the execution of our malicious application, likely due to its frequent access to connection state (e.g., repeated calls to \texttt{GetExtendedTcp\-Table}) in order to monitor active connections. SAC is enabled by default only on ``clean'' installations of Windows 11, and is disabled on systems upgraded from Windows 10. We therefore assume SAC is disabled in our evaluation, as a significant portion of Windows 11 installations are upgrades from Windows 10, and moreover, Windows 10 still has a significant market share.

An 
IP-options-based pre-processing technique was used, as described in \autoref{sec:isn_attack_windows}. 
Preprocessing for the Windows ephemeral port range (49152--65535 \cite{microsoft_dynamic_port_range}) took 1300 seconds 
and generated approximately 1.53 GB of network traffic 
(this is significantly more time and network traffic than Linux, due to the need to find correlative port to each target port, which involves sending hundreds of packets per target port). 

Source port detection was performed using the Windows equivalent of \texttt{/proc}-based monitoring, as described in \autoref{sec:os-level-connection-monitoring}. The ARN was placed within the local network. 
A simple Python client application using standard sockets was used as the TCP client.

To further validate the correlative ports behavior introduced in \autoref{sec:isn_windows}, we conducted a pairwise measurement experiment across two  
distinct machines: a Dell mini PC (i7-13700T) running Windows 11 Home (version 25H2) and a Latitude 5540 laptop running Windows 11 Pro (version 25H2). Specifically, we selected two separate sets of $200$ candidate source ports and evaluated every possible pair $(p_1, p_2)$ between them. For each pair, we rapidly initiated TCP connections using $p_1$ and $p_2$ as the source ports---while keeping the source IP, destination IP, and destination port constant---captured their respective ISNs, and calculated the ISN difference. If this difference remained stable across multiple trials, the pair was classified as correlative. Our empirical results show that for any given port $p_1$ in the first set, there were on average $49.5$ ($\sigma=0.75$) correlative ports out of the $200$ candidates in the second set on Machine A, and $48.89$ ($\sigma=1.54$) correlative ports on Machine B. These findings consistently substantiate our claim that approximately $25\%$ of candidate ports are correlative.

\ifextended
\paragraph{IP Options Implementation on Windows} 
On Windows, we observe a different behavior when using IP source routing options. Specifically, while the destination IP address is modified according to the next hop specified in the option, the destination layer 2 address (destination MAC address) is still set to the next hop in the routing path for the original destination IP. As a result, if a packet is sent to an external IP address (e.g., 192.0.2.123) 
with a source-routing option that redirects it to a host in the LAN, the packet will carry the LAN host's IP address as its IP destination address, but the layer 2 destination address will be set to the gateway (router) MAC address. Consequently, the packet is first delivered to the router, which must support and correctly process IP options in order to forward it to the host in the local network. Therefore, in our experiment, the attacker is located within the local network, and we use a Linux router that supports IP options 
(\texttt{net.ipv4.conf.all.accept\_source\_route=1}).
\fi

\begin{table*}[h]
\centering
\caption{TCP hijacking results across platforms. \normalfont ARN denotes the attacker remote node location; in our experiments, it may be split into capturing ISN and injecting ARN components. B.F. denotes the brute-force search space for the ISN. Each setting was tested 100 times on Linux and macOS, 500 times on Windows, 50 times on Android, and 25 times on iOS.} 
\begin{minipage}{0.95\textwidth}
\centering
\begin{tabular}{c c c c c c c c}
\toprule
OS & Scenario & IP & Capturing ARN & Injecting ARN  & B.F. & Success &
Local Source-IP Rewriting\\
\midrule
Linux & cBPF & IPv4 & N/A\footnote{N/A indicates that no capturing ARN was required for this attack.}& AWS & 1.6K & 97\%\footnote{In this experiment, the victim application was a simple Python client. We also evaluated Firefox as the victim client under otherwise identical conditions and achieved 8/10 successful attempts.}& Used\\
Linux & cBPF & IPv6 & N/A & AWS & 1K & 97\% & Used \\
Linux & IP Opt. & IPv4 & LAN & AWS  & 1.6K & 99\% & Used \\
Android & cBPF & IPv4 & N/A & AWS & 4.5K & 80\% & Used\\
Android & IP Opt. & IPv4 & LAN & AWS & 4.5K & 86\% & Used\\
macOS & IP Opt. & IPv4 & LAN & LAN  & 8K & 100\% & Not Used\\
iOS & IP Opt. & IPv4 & LAN & LAN & 6K &96\% & Not Used\\
Windows & IP Opt. & IPv4 & LAN & LAN  & 12K & 87\% & Not Used\\
\bottomrule
\end{tabular}
\end{minipage}

\begin{minipage}{0.95\textwidth}

\end{minipage}

\label{tab:tcp_hijack_results}
\end{table*}

\subsection{Setup for DNS Cache Poisoning --- Linux}
On Linux, source port detection was performed using three different methods: the \texttt{/proc}-based technique described in \autoref{sec:os-level-connection-monitoring}, \texttt{bind()}-based probing described in \autoref{sec:bind-port-scanning}, and ephemeral port sampling via \texttt{bind(0)} described in \autoref{sec:bind-zero-port-sampling}. Our malicious code was executed under one user account, after which the stub resolver cache was poisoned. We then verified that a second user account was affected, as both users shared the same system-wide DNS cache.
In this setup, the victim was hosted on a Linux Azure virtual machine, while the attacker operated from an AWS EC2 Linux instance.

We used a relatively low transmission rate of approximately $130\mathrm{K}$ packets per second. This rate can be further increased to achieve faster attacks. The results of a single round 
DNS cache poisoning attack against Linux with various source port detection methods are presented in \autoref{tab:cache_poisoning_Linux}. 

\subsection{Setup for DNS Cache Poisoning --- Android}
On Android, source port detection was performed using
\texttt{bind()}-based probing, as described in
\autoref{sec:bind-port-scanning}. Before each attack attempt, the
malicious application evicted the legitimate DNS record by issuing
640 queries \cite{aosp_dns_cache} to different domains, corresponding to the default size of the Android DNS
cache. Cache eviction required approximately
495 milliseconds.
As explained in \autoref{subsec:android-dns}, the effective race window is defined by $\text{RTT}_{\text{victim}\leftrightarrow\text{resolver}} - \text{RTT}_{\text{victim}\leftrightarrow\text{ARN}}$. In our case, $\text{RTT}_{\text{victim}\leftrightarrow\text{resolver}}$ (the RTT to OpenDNS, the resolver used in the experiment) was $60\,\mathrm{ms}$ and $\text{RTT}_{\text{victim}\leftrightarrow\text{ARN}}$ was $14\,\mathrm{ms}$, yielding an effective window of $46\,\mathrm{ms}$.

During this interval,  the
malicious application scanned 1,500 candidate UDP ports, which required
approximately 19.5 milliseconds.\footnote{\label{footnote-less-than-half}A bit shorter than half the attack window due to implementation constraints.} The ARN then used the remaining time to transmit 8000 forged DNS
responses with candidate TXIDs. This process is repeated until one of the forged responses is accepted.

The malicious code was implemented as a regular unprivileged Android
application. After a successful attack, we verified that the DNS cache
had been poisoned by resolving the target domain from a separate
application using \texttt{ping}. The experimental results are presented
in \autoref{tab:cache_poisoning_windows_android}. 
We explicitly calculate the time for a na\"{\i}ve brute-force attack under the same experimental conditions and show that it is completely infeasible:
with a transmission rate of $8{,}000$ packets in $22\,\mathrm{ms}$\footref{footnote-less-than-half} (roughly $364K$ 
packets per second) and an effective race window $T = 46\,\mathrm{ms}$, a single naive attempt can transmit at most $46\,\mathrm{ms} \times 364K\,\text{packets/sec} \approx 17K$ packets. 

Given that the total search space for a na\"{\i}ve attack (full Android ephemeral port range $\times$ 16-bit TXID) is $(65{,}534 - 1025 + 1) \cdot 65{,}536 \approx 4.23 \times 10^9$ combinations, the success probability per attempt is:
$$P_{\text{naive}} = \frac{17K}{(65534-1025+1)\cdot 65536} \approx 4 \times 10^{-6}$$

Using the geometric distribution, the expected number of attempts required to succeed is $E[N] = \frac{1}{P_{\text{naive}}} \approx 250K$ attempts. Each attempt requires evicting the entire cache ($495\,\mathrm{ms}$ on average), and adding the attack window itself results in roughly $555\,\mathrm{ms}$ in total per attempt. Thus the total expected time for the naive brute-force attack is:
$$250K \times 0.555\,\mathrm{seconds} \approx 139K\,\mathrm{seconds} \approx 38.6\text{ hours}$$

\subsection{Setup for DNS Cache Poisoning --- Windows}
On Windows, the resolver's UDP source port was identified using
\texttt{GetExtendedUdpTable}, as described in
\autoref{sec:os-level-connection-monitoring}. Before each attack
attempt, the malicious application flushed the DNS cache using
\texttt{DnsFlushResolverCache}. Cache flushing required approximately
$140\,\mathrm{\mu s}$.

As discussed earlier, the attacker can use only one TXID guess per query
if the forged response is to be cached. To shorten failed attempts, we
sent the same TXID guess five times. If the guess was correct, the
forged response would be accepted and cached. If it was incorrect, the resolver would terminate the query, allowing the attacker to proceed more quickly to the next attempt.

In our experiments using the OpenDNS resolver, the network latencies were $\text{RTT}_{\text{victim}\leftrightarrow\text{resolver}} = 51\,\mathrm{ms}$ and $\text{RTT}_{\text{victim}\leftrightarrow\text{ARN}} = 5\,\mathrm{ms}$. Because the stub resolver terminates the query upon receiving five packets with an incorrect TXID, our technique of repeating the guess forces an early timeout, reducing the duration of a single round from approximately $55\,\mathrm{ms}$ to just $7.4\,\mathrm{ms}$. Furthermore, since only five packets are transmitted per attempt, the transmission overhead incurred by the ARN is negligible.

Windows maintains separate DNS caches for \texttt{AF\_INET} and
\texttt{AF\_UNSPEC} queries~\cite{klein_ndss_2019}.
\texttt{AF\_INET} requests IPv4 addresses only, whereas
\texttt{AF\_UNSPEC} allows either IPv4 or IPv6. In our experiments, we
targeted the \texttt{AF\_INET} cache. 

The malicious code was executed as a regular unprivileged Windows
application. After a successful attack, we verified the poisoned entry
by resolving the target domain using \texttt{ping} from another user
account. The experimental results are presented in
\autoref{tab:cache_poisoning_windows_android}.
\ifextended
When compared to a na\"{\i}ve brute-force approach, an attacker must similarly constrain themselves to one TXID guess per attempt. Lacking knowledge of the exact source port, a na\"{\i}ve attacker can attempt to spray a single TXID guess across the entire Windows ephemeral port range (ports 49152--65535, totaling $16{,}384$ ports).  Given a resolver RTT of approximately $50\,\mathrm{ms}$, as in our experiment, the effective race window is sufficiently large to allow the transmission of all $16{,}384$ packets before the genuine response arrives. 

Crucially, however, because this na\"{\i}ve strategy delivers only a single packet to the correct port, it fails to trigger the resolver's five-packet early termination defense. As a result, each na\"{\i}ve attempt incurs the full $50\,\mathrm{ms}$ resolution time plus the $140\,\mathrm{\mu s}$ cache flush, taking roughly 
$50.14\,\mathrm{ms}$ per round. Since the success probability per attempt is dictated purely by the TXID guess ($P_{\text{naive}} = \frac{1}{65536}$), the expected number of attempts required to succeed is $E[N] = 65{,}536$. The total expected time for the na\"{\i}ve brute-force attack is therefore:
$$65{,}536 \times 0.05014\,\mathrm{seconds} \approx 3286\,\mathrm{seconds} \approx 54.76\text{ minutes}$$

While this makes the na\"{\i}ve attack feasible under a $50\,\mathrm{ms}$ RTT, it requires massive network overhead ($16{,}384$ packets per attempt compared to our stealthy $5$ packets) and takes substantially longer: theoretically $3286$ seconds, versus an average of $515$ seconds for our attack, as shown in \autoref{tab:cache_poisoning_windows_android}.
\fi

\begin{table}[h]
\centering
\caption{DNS cache poisoning results on Linux. \normalfont Results are based on 100 experiments 
for procfs and bind-all, and 50 experiments for \texttt{bind(0)}. 
Local source-port rewriting was used. 
Each experiment consisted of a single round only (inferring a port and brute-forcing the TXID).
}
\label{tab:cache_poisoning_Linux}
\begin{tabular}{ c c c c}
\hline
Port Detection & Success Rate & Packets & Max Exp. Time \\
\hline
procfs   & 97\% &  $65\mathrm{K}$ & $0.65\mathrm{s}$  \\ 
\texttt{bind()}-all &  99\% & $2 \times 65\mathrm{K}$ & $1.2\mathrm{s}$  \\
\texttt{bind(0)}  &  68\% & $2 \times 65\mathrm{K}$  & $2.85\mathrm{s}$\\
\hline
\end{tabular}
\end{table}

\begin{table}[h]
\centering
\caption{DNS cache poisoning results on Windows and Android. \normalfont Results are based on 10 experiments. Local source-port rewriting was used. 
}

\label{tab:cache_poisoning_windows_android}

\begin{tabular}{c c c c}
\hline
OS 
& \shortstack{Success\\Rate}
& \shortstack{Packets\\(mean; $\sigma$)}
& \shortstack{Exp. Time\\(mean; $\sigma$)} \\
\hline

Windows & $100\%$ &
$\left(\mathbf{71K}; \sigma=\mathrm{108K}\right)\mathbf{\times} \mathbf{5}$ &
$\mathbf{515}\mathrm{s}; \sigma=753\mathrm{s}$ \\

Android & $100\%$ &
$\left(\mathbf{10}; \sigma=9.46\right)\mathbf{\times 8\mathrm{K}}$ &
$\mathbf{471}\mathrm{s}; \sigma=419\mathrm{s}$ \\

\hline
\end{tabular}
\end{table}

\section{Discussion}
In addition to the attacks described above, our experiments revealed two broader implications: a potential timing side channel in system calls and privacy implications of TCP ephemeral-port selection. We discuss these observations in \reffulldiscussion. We next focus on practical mitigations for the vulnerabilities demonstrated in this work.

\subsection{Attack Mitigation}

In this subsection, we outline the key security weaknesses identified in our work and propose practical mitigations.

\begin{enumerate}
\item \textbf{Information Leakage via System Interfaces.}  
Unprivileged applications can infer the connection state of other processes. In particular, interfaces such as \texttt{/proc/net} expose information about active connections (e.g., port usage), enabling cross-user information leakage. We see no strong justification for exposing such data to unprivileged applications. A straightforward mitigation is to preserve access to these interfaces while filtering their output on a per-user basis, so that each user can view only sockets owned by that same user. Notably, modern Android systems already enforce stronger (but less granular) restrictions, preventing regular applications from accessing \texttt{/proc/net}.

\ifextended
\item \textbf{Port Inference via Binding Behavior.}  
Another leakage channel arises from the ability to probe port allocation. By repeatedly invoking \texttt{bind()}, an application can infer which ports are currently in use.  
Completely preventing this behavior is impractical, as legitimate applications rely on socket operations. Instead, we propose rate-limiting such operations. In particular, the number of \texttt{bind()} and \texttt{connect()} calls should be limited per user (rather than per process), to prevent bypass via process or thread spawning. Privileged processes (e.g., those with \texttt{CAP\_NET\_BIND\_SERVICE}) may be exempted. This can be enforced separately per transport protocol (UDP and/or TCP). 

\item \textbf{Port Exhaustion and Allocation Manipulation.}  
Attackers can exhaust the ephemeral port space to force predictable allocation, as shown by Alharbi et al.~\cite{collaborative_client_side_dns} and Hay et al.~\cite{dns_posioning_via_port_exhaustion_2011}.
We propose partitioning the ephemeral port range into separate regions for privileged and unprivileged applications. For example, the upper half of the range can be reserved for privileged processes, while unprivileged applications are restricted to the lower half.  
This prevents both port exhaustion and port inference attacks targeting privileged services, as well as attacks that rely on continuously binding ports to detect which ports are occupied. A simpler mitigation that addresses only port-exhaustion attacks is to enforce a per-user limit on the total number of simultaneously allocated ports for both TCP and UDP sockets, thereby reducing the feasibility of such attacks.

\item \textbf{Predictable Ephemeral Port Allocation.}  
The Linux 
TCP ephemeral port selection algorithm introduces predictability: consecutive connections to the same destination use ports that differ by small increments.  
We recommend increasing the entropy of port selection by removing deterministic increments and strengthening randomization.
\fi

\item \textbf{Predictable ISN Generation.}  
The ISN generation algorithm remains partially predictable for a fixed 4-tuple. We propose incorporating additional context into the computation, such as the user identifier (UID), 
as in $\text{ISN} = H_k(\textit{4-tuple}, \text{UID}) + \mathit{time} \mod 2^{32}$.
Including the UID ensures that ISNs generated by different users are no longer correlated, even when other connection parameters are identical. This prevents unprivileged applications from inferring or predicting sequence numbers of connections initiated by other users, thereby strengthening isolation and mitigating cross-user side-channel attacks.

\item \textbf{cBPF Leakage Channel.}  
The cBPF mechanism allows filters to control how many bytes are delivered to user space based on packet contents, enabling leakage of sensitive fields such as TCP sequence numbers. This can be mitigated by sanitizing packet data before exposure, for example by zeroing out sequence and acknowledgment numbers (and zeroing/adjusting the TCP checksum accordingly). Alternatively, systems may disable the use of cBPF-based filtering on TCP sockets for non-privileged applications to prevent such leakage altogether.

\item \textbf{IP Options Abuse.}  
Certain IP options, specifically Loose Source and Record Route (LSRR) and Strict Source and Record Route (SSRR), allow attackers to influence packet paths and observe traffic at chosen locations. 
We reinforce existing recommendations to block the use of these source-routing options in TCP traffic across the Internet~\cite{rfc7126}.

In particular, end hosts should require elevated privileges to send TCP packets with source-routing IP options, specifically LSRR and SSRR.
\end{enumerate}

Additional potential mitigations are presented in the extended paper \cite[Section 9.1]{extended-paper}.

\section{Conclusion}

We 
take a fresh look at the attack model in which a low-privilege malicious application, co-located on the victim host, collaborates with a remote adversary to compromise network protocols. Within this model, we showed that core defenses in TCP and DNS can be bypassed via residual information leakage from operating system interfaces and network stack behavior.

We developed techniques for inferring sensitive connection state, including TCP ISN and DNS source port, using a malicious app/user on the target machine 
without requiring privileged access or packet capturing capabilities. Using these primitives, we demonstrated practical TCP connection hijacking and DNS cache poisoning attacks against other apps/users on the machine, 
and evaluated them across Linux, Android, macOS, iOS, and Windows. Overall, our findings indicate that core leakage primitives persist across platforms despite differences in implementation, even though certain mechanisms (cBPF-based leakage) are specific to particular operating systems.
These findings challenge common assumptions (which are implied in RFC 6528 and RFC 9293) about the traditional off-path attacker model by demonstrating that limited local code execution  which is still off-path 
suffices to violate fundamental security 
properties. This calls for rethinking network protocol security under adversaries with local (even unprivileged) footholds, with an emphasis on stronger isolation and elimination of cross-application leakage channels.

\begin{acks}
We are grateful to the anonymous reviewers and shepherd of ACM CCS 2026 for their thoughtful reviews and feedback. 
We thank Inbal Schussheim for her insightful observations regarding DNS cache poisoning on macOS and iOS, which helped improve this work. We thank Inbal Schussheim, Oriyan Hermoni and Noam Caspi for their useful feedback on an earlier draft of this manuscript. This research was partially funded by  ISF grant 1071/23.
\end{acks}

\bibliographystyle{ACM-Reference-Format}
\balance
\IfFileExists{ref.bbl}{

}
{ 
\bibliography{ref}
}

\appendix

\section{Open Science}
An artifact accompanying this work is available at: \url{https://github.com/TamirShahar/artifacts/}.
The artifact supports the reproduction of three attacks on Linux systems. 
\ifextended
We demonstrate TCP session hijacking via ISN prediction, where a malicious unprivileged application performs the inference stage and coordinates with an off-path attacker to complete the attack. We present two alternative pre-processing approaches for this inference: a cBPF-based technique and an IP options--based technique. In addition, we demonstrate cache poisoning of the Linux stub resolver (systemd-resolved), showing how a malicious unprivileged user can coordinate with an off-path attacker to poison the DNS cache.

All code and configuration required to reproduce the attacks are included. The environment is deployed using Docker containers interconnected via an on-host virtual network, with all entities residing within the same (virtual) LAN for demonstration purposes. It includes a victim host with two unprivileged users, an ARN with packet crafting capabilities, and auxiliary services such as an HTTP server, a DNS recursive resolver and a DNS authoritative nameserver required to demonstrate the cache poisoning attack.
The artifact is designed to run on a single machine with a standard 64-bit Linux environment running Docker Engine. Where relevant, software is locked to specific versions, in order to ensure consistent and reproducible behavior across different setups.

Detailed setup instructions and step-by-step guidance for reproducing each attack are provided in the accompanying README file. The artifact also includes scripts for executing both the pre-processing stages and the full attacks, as well as instructions for evaluating and validating the experimental results.

\fi
In addition, we provide details about our reverse-engineering methodology and artifacts in \refreveng.

\section{Ethical Considerations}
\ifextended
\subsection{Disclosures}
\fi
We disclosed the vulnerabilities to the OS vendors, including Linux maintainers, Apple, Microsoft and Google, beginning on April 29\textsuperscript{th}, 2026, providing technical details, and suggested mitigations. We received responses from all vendors, with different remediation outcomes. The Linux maintainers have already deployed two patches (CVE-2026-\anonymize{XXXX}{53236} and CVE-2026-\anonymize{YYYY}{53249}) addressing the TCP hijacking attack, covering the cBPF-based and IP-options-based leakage mechanisms. For the Linux DNS cache poisoning issue, the maintainers characterized it as an inevitable limitation of shared systems and did not plan a patch. The reports on the TCP attack remain under review by Microsoft and Apple. Google classified the Android TCP hijacking issue as Low severity and determined that the Android DNS cache poisoning issue does not constitute a security vulnerability; both were logged for potential future remediation. Microsoft does not plan to fix the DNS issue, stating that customers should use DoH or DoT.
\ifextended
The current status of the disclosures is summarized in \autoref{tab:disclosure}. 
\begin{table*}[t]
\centering
\caption{Coordinated disclosure status 
}
\label{tab:disclosure}
\begin{tabular}{l l l l p{6.2cm}}
\toprule
\textbf{Vendor} &
\textbf{Platform} &
\textbf{Attack} &
\textbf{Report Date} &
\textbf{Patch/Mitigation Status} \\
\midrule

Linux maintainers
& Linux
& TCP hijacking
& April 29\textsuperscript{th}, 2026
& Patched. The Linux maintainers directly mitigated the cBPF-based leakage mechanism (CVE-2026-\anonymize{XXXX}{53236}) and the IP-options-based mechanism (CVE-2026-\anonymize{YYYY}{53249}).
\\

Linux maintainers
& Linux
& DNS cache poisoning
& April 29\textsuperscript{th}, 2026
& 
Acknowledged; no patch is currently planned. The Linux maintainers describe this attack as ``inevitable limitations of shared systems'' and consider it less critical.

\\

Google
& Android
& TCP hijacking
& June 23\textsuperscript{rd}, 2026
& Meanwhile, it was fixed by Linux. \\

Google
& Android
& DNS cache poisoning
& July 1\textsuperscript{st}, 2026
& Determined not to be a security vulnerability; logged for potential future remediation. \\

Apple
& macOS/iOS
& TCP hijacking
& April 29\textsuperscript{th}, 2026
& Last update from Apple (September 3\textsuperscript{rd}, 2026) is that they are working on a fix based on restricting the source routing IP option. \\

Microsoft
& Windows
& TCP hijacking
& April 29\textsuperscript{th}, 2026
& Last update from Microsoft (August 7\textsuperscript{th}, 2026) is that they assess the report. Numerous requests for update/clarification since were not answered.\\

Microsoft
& Windows
& DNS cache poisoning
& July 1\textsuperscript{st}, 2026
& Microsoft stated that this falls outside their threat model, and concluded that ``to protect against techniques like this, we [Microsoft] recommend using DNS over HTTPS (DoH) or DNS over TLS (DoT)''. \\

\bottomrule
\end{tabular}
\end{table*}

\subsection{Stakeholder Analysis}
We discuss the stakeholders potentially affected by the vulnerabilities described in this work and the implications of their disclosure.

\begin{itemize}

\item \textbf{OS Vendors.}
We disclosed our findings to the security teams at Microsoft, Apple, Linux and Google, 
giving them 60 days to prepare a fix. This is in line with standard responsible disclosure procedures. We expect these vendors to produce and deploy their fix as part of their security release cycles. We plan to release our paper no earlier than 60 days \emph{after} the security patches become available, to provide customers of these product enough time to update their operating systems.

\item \textbf{Client Systems (End Users).}
End users running TCP-based applications and stub resolvers may be affected by the demonstrated attacks. However, most modern systems receive automatic security updates. We coordinate our disclosure to allow sufficient time between patch availability and publication (60 days), thereby drastically reducing their exposure window.

\item \textbf{Domain Owners.}
Domain owners are not directly affected. The demonstrated DNS cache poisoning attacks do not exploit flaws in resolvers or stub resolvers, but rather OS-level behavior. Once the operating system patches are publicly available and deployed, the attack is expected to no longer be feasible.

\item \textbf{Application Servers.}
Application servers are not directly compromised but may be temporarily impersonated in TCP hijacking attacks. During the attack, the victim may generate inconsistent TCP packets (e.g., out-of-window acknowledgments or unexpected resets), which can reach the server and appear as anomalous traffic. However, since the root cause lies in client-side OS behavior, the risk is mitigated once patches are deployed, and normal operation is expected to be restored.

\end{itemize}

\fi

\section{Generative AI Usage}

We used generative AI tools, including ChatGPT and GitHub Copilot, during the preparation of this work. They were used to assist with drafting and refining portions of the text, including grammar, spelling, and phrasing. In some cases, they were also used to help summarize background information on specific mechanisms discussed in the paper. All such content was carefully reviewed, verified against official documentation and prior literature, and, where applicable, validated through our own experiments. In addition, AI tools were used to generate code skeletons. All generated code was manually inspected, with function usage verified against official documentation and external sources. We further validated the correctness of the code by running controlled experiments and ensuring that each component behaved as intended.

\ifextended
\clearpage
\begin{table*}[t]
\section{Comparison with Prior Work}
\label{app:prior_work_comparison}

\small
\setlength{\tabcolsep}{4pt}
\centering
\begin{tabular}{p{3.5cm} @{\hspace{0.5cm}} p{5.0cm} p{8.0cm}}
\toprule
\textbf{Prior work} &
\textbf{Main approach} &
\textbf{Difference from our work} \\
\midrule

\multicolumn{3}{c}{\textbf{\textit{Host-Based and OS-Level Side Channels}}} \\
\midrule

Qian et al. (2012)~\cite{collaborative_tcp_squence_number_inference}
&
Collaborative TCP sequence-number inference using local unprivileged code
&
Same 
threat model. Their techniques for Android, iOS, and macOS ---which relied on globally available TCP counters --- are already mitigated, 
no Windows attack was demonstrated, the Linux approach may exhaust TCP ports, and DNS poisoning was not considered. \\

Chen et al. (2015)~\cite{static_detectino_of_packet_injection_vulnerabilities}
&
Static detection of packet-injection vulnerabilities
&
Focuses on vulnerability detection rather than end-to-end exploitation and depends on Qian et al.'s techniques for Linux attacks. \\

Zhang et al. (2018)~\cite{leakage_on_ios}
&
OS-level side channels for cross-application activity inference on iOS
&
Infers application activity; we use leaked network state to enable practical TCP hijacking. \\

Spreitzer et al. (2018)~\cite{procHarvester}
&
procfs 
side channels for cross-process activity inference
&
Focuses on activity leakage; we infer TCP and DNS connection state, including on platforms such as Android where relevant \texttt{procfs} access is restricted. \\

\midrule
\multicolumn{3}{c}{\textbf{\textit{Off-Path TCP Injection and State Inference}}} \\
\midrule

Gilad et al. (2012)~\cite{attacking_the_web}
&
Global IPID counters for TCP state inference and injection
&
Relies on globally shared IP ID counters that were later modified; we do not rely on the IP ID implementation. \\

Cao et al. (2016)~\cite{global_rate_limit}
&
Global challenge-ACK rate-limit side channel
&
The side channel was mitigated; our techniques do not rely on the global challenge-ACK counter. \\

Feng et al. (2020)~\cite{mixed_ipid_assigment} --- Mixed IPID
&
Mixed IPID assignment for TCP state inference
&
This behavior was subsequently fixed; our attacks use different OS-level primitives. \\

Feng et al. (2022)~\cite{PMTUD_is_not_Panacea} --- Fragmentation
&
IPv4 fragmentation and IPID guessing
&
Depends on IPv4 fragmentation; our TCP attack does not require fragmentation and also applies to IPv6. \\

Wang et al. (2024)~\cite{packet_size_side_channel} ; Li et al. (2025)~\cite{packet_length_leakage}
&
Packet-size and packet-length leakage in Wi-Fi or cellular networks
&
Require wireless-network visibility; our attack does not depend on wireless-specific characteristics. \\

Feng et al. (2025)~\cite{pmtud_breaks_tcp} --- PMTUD
&
Shared PMTUD state and a local-network puppet
&
Requires a controlled local-network machine with fine-grained observations; we require only an unprivileged application on the victim. \\

Yang et al. (2026)~\cite{session_manipulation_attacks}
&
VPN connection-tracking side channels
&
Requires VPN-specific deployment conditions; our techniques do not depend on a VPN. \\

\midrule
\multicolumn{3}{c}{\textbf{\textit{DNS Cache Poisoning Attacks}}} \\
\midrule

Alharbi et al. (2019)~\cite{collaborative_client_side_dns}; Hay et al. (2011)~\cite{dns_posioning_via_port_exhaustion_2011}
&
UDP-port exhaustion for DNS cache poisoning
&
Exhausts the port space and may disrupt applications; we infer the resolver port without port exhaustion. \\

Herzberg and Schulmann (2012)~\cite{fragmentation_considered_poisonous}
&
DNS fragmentation attacks
&
Relies on IP fragmentation (which current best practices recommend avoiding, e.g., by rejecting fragmented DNS traffic) and primarily targets recursive resolvers or forwarders; we target the local stub resolver without fragmentation. \\

Klein (2021)~\cite{cross_layers_attack_klein}
&
Prediction of UDP source ports from weak kernel randomness
&
Exploits a randomness weakness that was fixed; we infer the actual resolver source port through local OS interaction. \\

Gierlings et al. (2023)~\cite{Isolated_and_exhausted}
&
Browser-based UDP-port exhaustion
&
Requires browser-mediated exhaustion and was mitigated in modern browsers; our attack uses a local unprivileged application. \\

\midrule
\textbf{This work}
&
\textbf{New OS-level primitives for ISN and source-port inference}
&
\textbf{Supports contemporary Linux, Android, Windows, macOS, and iOS, and demonstrates both TCP hijacking and DNS cache poisoning under one collaborative attacker model.} \\

\bottomrule
\end{tabular}
\end{table*}

\clearpage
\section{Analysis of Ephemeral Port Sampling via \texttt{bind(0)}}
\label{app:bind-zero-analysis}

In Method~5 (\autoref{sec:bind-zero-port-sampling}), we rely on repeated invocations of \texttt{bind(0)} to sample ephemeral ports assigned by the kernel. This raises the question of whether such sampling can reliably cover the set of available ports.

Let $L$ denote the number of available ephemeral ports, and let $N$ be the number of independent \texttt{bind(0)} invocations. Each invocation returns a port selected according to the kernel's internal (pseudo-randomized) allocation policy. While the selection is not guaranteed to be uniform, we approximate it as such for the purpose of analysis.

\paragraph{Single-port coverage.}
The probability that a specific port is \emph{not} selected in a single trial is $(1 - \frac{1}{L})$. Over $N$ independent trials, the probability that it is never selected is:
\[
\left(1 - \frac{1}{L}\right)^N.
\]
For large $L$, we use the approximation $(1 - \frac{1}{L})^L \approx e^{-1}$, yielding:
\[
\left(1 - \frac{1}{L}\right)^N \approx e^{-N/L}.
\]
Therefore, the probability that a given port \emph{is} sampled at least once is:
\[
1 - e^{-N/L}.
\]

\paragraph{Full coverage.}
Assuming independence across ports (an approximation), the probability that all $L$ ports are sampled at least once is:
\[
\left(1 - e^{-N/L}\right)^L.
\]
Assuming $N \gg L$ (as we shall see below), and using the approximation $(1 - x) \approx e^{-x}$ for small $x$, and since $e^{-N/L} \ll 1$ (as $N \gg L$), we obtain:
\[
\left(1 - e^{-N/L}\right)^L \approx e^{-L \cdot e^{-N/L}}.
\]

\paragraph{Choosing $N$.}
Let $N = \alpha L \log L$ for some $\alpha > 1$. Substituting:
\[
e^{-N/L} = e^{-\alpha \log L} = L^{-\alpha},
\]
and thus the probability of full coverage becomes:
\[
e^{-L \cdot L^{-\alpha}} = e^{-L^{1-\alpha}}.
\]
For $\alpha > 1$ and large $L$, this can be approximated as:
\[
1 - \frac{1}{L^{\alpha - 1}},
\]
which approaches 1 rapidly as $L$ grows, especially for large values of $\alpha$.

\paragraph{Implications.}
This analysis shows that although \texttt{bind(0)} sampling does not guarantee coverage of all available ports, it achieves near-complete coverage with high probability after $O(L \log L)$ samples. In practice, this means that ports that are never observed during sampling are \emph{likely} to be in use, though false negatives remain possible due to randomness in the allocation process.

Importantly, this explains the limitation noted earlier: the absence of a port from the sampled set does not strictly imply that it is occupied, but only that it was not selected during the sampling process. Nevertheless, with sufficiently many samples, this ambiguity becomes negligible.

\section{Full Discussion} 
\label{app:full_discussion}
\subsection{Timing Side Channels in System Calls}
\label{app:time_side_channel}
In the context of source port discovery using the Linux ephemeral port selection algorithm described in \autoref{sec:linux-ephemeral-port-selection}, our malicious code repeatedly invoked \texttt{bind()} in a tight loop. While measuring the execution time of these calls, we observed occasional latency spikes that appeared to correlate with concurrent port allocations. A plausible explanation is contention on shared kernel resources, such as locks protecting data structures involved in port allocation, causing one system call to delay another. This suggests that syscall timing may expose information about concurrent activity within the kernel. While we did not systematically explore this effect, it points to a broader class of timing-based side channels that extend beyond networking. Prior work by Hund et al.~\cite{timing_sidechennel_aslr} showed that timing differences observable from unprivileged code can leak information about kernel memory layout (e.g., KASLR), and Gu et al.~\cite{timing_sidechannel_syncfs} demonstrated that syscall latency (e.g., via the \texttt{syncfs} system call) can be exploited to infer system-wide I/O activity. Unlike these works, we do not develop a full attack based on syscall timing, and leave the systematic exploration of such channels in the context of networking to future work.

\subsection{Privacy Implications of TCP Ephemeral Port Selection}
TCP ephemeral port selection can introduce privacy risks. 
For example, our Linux-specific port discovery technique \autoref{sec:linux-ephemeral-port-selection} which relies on the Linux port allocation algorithm, can be adapted to detect new TCP connections to a set of particular destination hosts and ports, from an unprivileged application (note that for this, there is no need for ARN). To leak information about application behavior, this port discovery technique can be employed to repeatedly test if the predicted port for a connection to an IP address and port of interest (or a small set thereof) is occupied. Once the port becomes occupied, it marks the establishment of a new TCP connection to this destination from a victim application.
Prior work by Spreitzer et al.~\cite{procHarvester} demonstrated website inference using information exposed via \texttt{/proc}, which has since been restricted on Android; our approach remains effective even under these restrictions. Kol et al.~\cite{285415} exploit Linux's TCP ephemeral port selection algorithm (RFC 6056's Algorithm 4) to fingerprint devices remotely via a browser, but they do not describe detecting connections of other applications.

\clearpage

\begin{figure*}[t]
\section{TCP Hijacking Attack Flow}
\label{app:tcp-hijacking-flow6_4}

\centering
\includegraphics[height=0.95\textheight, keepaspectratio]{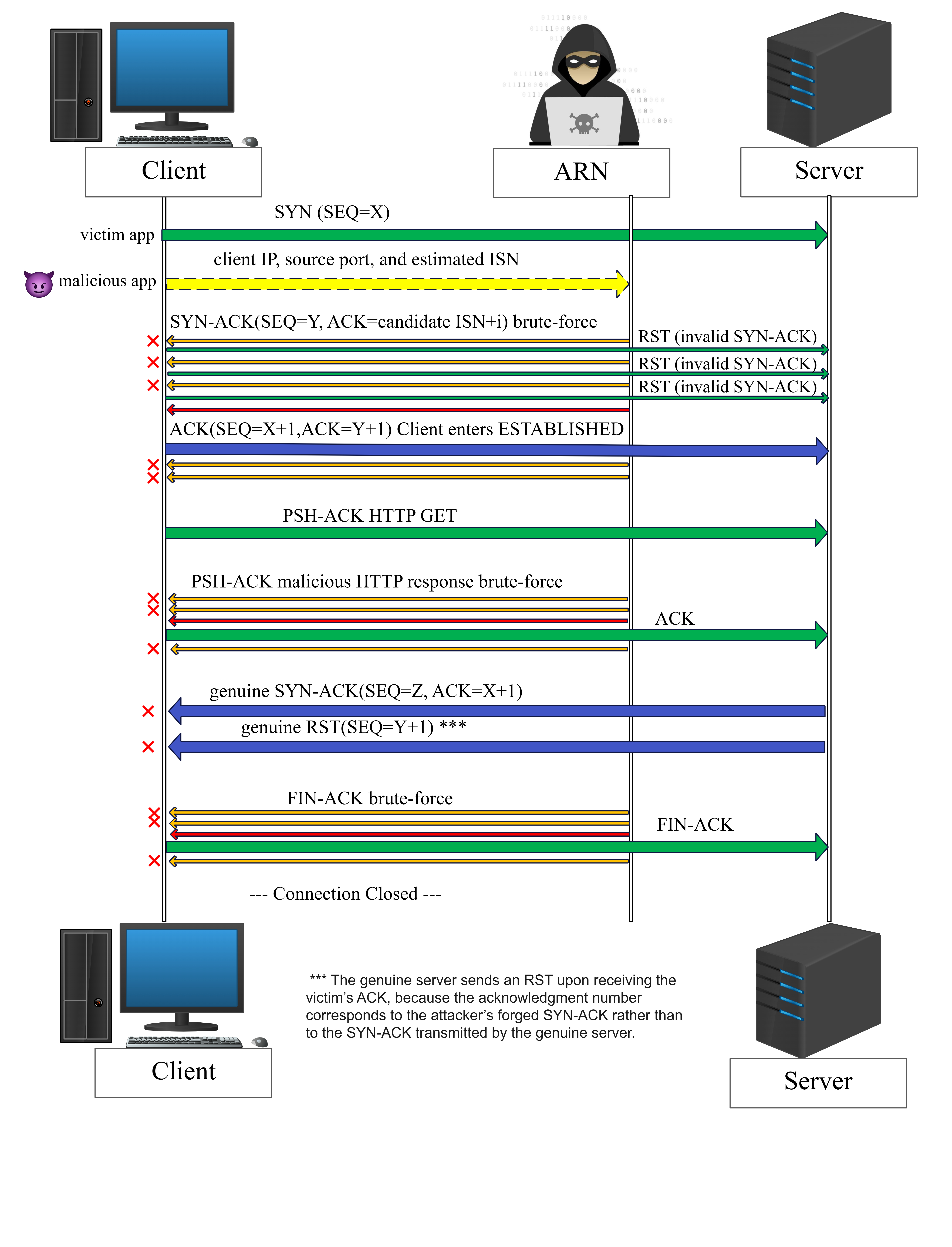}

\end{figure*}

\clearpage
\section{Reverse-Engineering Methodology and Artifacts}
\label{app:reverse-engineering}

This appendix describes the reverse-engineering methodology used to
analyze TCP ISN generation across the
examined operating systems. Our analysis combined source-code review,
static binary analysis, dynamic kernel debugging, and black-box network
experiments.

\subsection{Linux and Android}
We analyzed Linux kernel version 7.0.0-rc6.
We inspected the upstream Linux implementation of TCP ISN generation in \texttt{net/core/secure\_seq.c}~\cite{linux-secure-seq}.
Our analysis focused on the functions
\texttt{secure\_tcp\_seq\_and\_ts\_off()},
\texttt{secure\_tcpv6\_seq\_and\_ts\_off()}, and
\texttt{seq\_scale()}, and on the call paths through which TCP
connection parameters and timing information reach these functions.

We performed the analysis directly on the Linux kernel source code, following the relevant callers and data flow to identify the inputs used during ISN generation and the point at which the time-dependent component is incorporated. For Android, no separate reverse-engineering analysis was required, because the relevant TCP ISN generation logic is inherited from the
Linux kernel implementation.

\subsection{macOS and iOS}
We inspected the publicly available XNU implementation of TCP ISN
generation (version 12377.81.4). Our analysis focused on \texttt{tcp\_new\_isn()} in
\texttt{bsd/netinet/tcp\_subr.c} and on the helper functions and call
paths involved in constructing the initial sequence number
~\cite{xnu-tcp-subr}.
Because both macOS and iOS use the XNU networking stack, the same analysis
applies to both platforms.

\subsection{Windows}
Windows presented a more challenging case because its TCP/IP stack is
closed source. We therefore reverse engineered \texttt{tcpip.sys} version 10.0.19041.6157 (Windows 10), the
kernel driver implementing the Windows TCP/IP stack.

We assumed that the ISN-generation routine would invoke a cryptographic hash function as part of its computation. We therefore first located the implementation of the cryptographic hash routine by identifying references to \texttt{MD5ComputeTransform}, and then traced its callers until reaching \texttt{TcpCreateAndConnectTcbRateLimitComplete}, which participates in
TCP connection initialization and ISN generation (in newer versions of \texttt{tcpip.sys}, this logic has moved to \texttt{TcpCreateAndConnectTcbInspectConnectComplete}). 

The binary was analyzed statically using IDA Pro. We assigned
descriptive names to functions and variables as their roles became
clear. We complemented the static analysis with dynamic kernel
debugging using WinDbg, placing breakpoints in the relevant code paths
and observing the values used during TCP connection establishment.

Our analysis indicates that the Windows ISN is derived from a keyed
hash of the connection 4-tuple together with an additional evolving
offset. The offset appears to depend on time, recent connection
activity, and internal per-bucket state. However, we did not fully
recover all details of this mechanism.

We therefore complemented the reverse engineering with black-box
experiments. We repeatedly created pairs of TCP connections while keeping the source IP address, destination IP address, and destination port fixed and
varying the source ports. These experiments revealed that some
source-port pairs exhibit a stable relationship between their generated
ISNs when the connections are initiated within a short interval. We
refer to such pairs as \emph{correlative ports}.

The Windows analysis remains partial. In particular, although we
identified the relevant ISN-generation function and several components
of its internal state, additional work would be required to recover the
complete offset-update mechanism.

\subsection{Windows Artifacts}
The Windows reverse-engineering artifacts include the annotated IDA database for \texttt{tcpip.sys}, containing our function and variable naming, together with a document that walks through the main reverse-engineering observations and analysis. These
artifacts are 
available at:  
\url{https://github.com/TamirShahar/windows-reversing-artifacts/}
\fi

\end{document}